\documentclass[aps,prb,twocolumn,groupedaddress,floatfix]{revtex4}
\usepackage{graphicx}

\begin{document}

\title{Optical conductivity of a metal-insulator transition\\
for the Anderson-Hubbard model in 3 dimensions away from 1/2 filling}
\author{X. Chen}
\affiliation{Department of Physics, Queen's University, Kingston, ON K7L 3N6, Canada}
\author{R. J. Gooding}
\affiliation{Department of Physics, Queen's University, Kingston, ON K7L 3N6, Canada}

\date{\today}

\begin{abstract}
The Anderson-Hubbard model is considered to be the least complicated model using lattice fermions
with which one can hope to study the physics of transition metal oxides with spatial disorder. We
have completed a numerical investigation of this model for three-dimensional simple cubic 
lattices using a real-space self-consistent Hartree-Fock decoupling approximation for the Hubbard interaction.
In this formulation we treat the spatial disorder exactly, and therefore we {\em account
for effects arising from localization physics}.
We have examined the model for electronic densities well away 1/2 filling, thereby avoiding the physics
of a Mott insulator. Several recent studies have made clear that the combined effects of electronic 
interactions and spatial disorder can give rise to a suppression of the electronic density of states, 
and a subsequent metal-insulator transition can occur. We supplement such studies by calculating
the ac conductivity for such systems. Our numerical results show that
weak interactions enhance the density of states at the Fermi level 
and the low-frequency conductivity, there are no local magnetic moments, and the ac conductivity is Drude-like. 
However, with a large enough disorder strength and larger interactions the density of 
states at the Fermi level and the low-frequency conductivity 
are both suppressed, the conductivity becomes non-Drude-like, and these phenomena are accompanied by the 
presence of local magnetic moments. The low-frequency conductivity changes from a
$\sigma - \sigma_{dc}\sim\omega^{1/2}$ behaviour in the metallic phase, to a $\sigma\sim\omega^2$
behaviour in the nonmetallic regime. For intermediate disorder at 1/4 electronic filling, a metal-to-insulator 
transition is predicted to take place at a critical $U/B \approx 0.75$ ($U$ being the Hubbard interaction 
strength and $B$ the electronic band width). \textit{Our numerical results show that 
the formation of magnetic moments is essential to the suppression of the density of states at the
Fermi level, and therefore essential to the metal-insulator transition.} At weaker disorder a small lessening of
the density of states at the Fermi level occurs, but screening suppresses the spatial disorder 
and with increasing interactions no metal-insulator transition is found. 
\end{abstract}


\maketitle

\section{Introduction}
Electron-electron interactions are important in understanding the properties of many transition metal 
oxides.\cite{CoxTMO,Imada_MIT}  Novel ordered phases are found in this class of compounds: e.g., high-$T_c$ 
superconductors, $d$-electron heavy fermions, and quantum magnets. The inclusion of disorder into
such materials, and into the models of such systems, adds another level of complexity. In this report 
we focus on metal-to-insulator transitions that are controlled by the combined effects of both 
interactions and disorder. In part, this work is part of an effort to understand several experimental results,
including the properties of the weakly doped cuprates,\cite{Lai_PRB,Kastner_RMP} for which
disorder effects are known to be important. 

Initially, we were motivated to conduct such theoretical work to better understand this transition in the 
material LiAl$_y$Ti$_{2-y}$O$_4$.\cite{Lambert90} The ordered and undoped material LiTi$_2$O$_4$ 
undergoes a transition to a superconducting phase\cite{Johnston76I,Johnston76II} around $T_c\approx 12~K$, 
and has been proposed to be related to the family of high-$T_c$ superconductors.\cite{Mueller96}
If such a conjecture is correct, there must be reasonably large electron-electron interactions in this system.
The Al-doped system undergoes a metal-insulator transition for $y\approx 0.33$, and recent theoretical 
work by one of us and co-workers have
suggested that (i) the effects of disorder alone cannot lead to such a transition\cite{Fazileh_PRB};
(ii) inclusion of electron-electron interactions in the form of an on-site Hubbard energy can lead 
to such a transition.\cite{Fazileh_PRL} 
This occurs by the suppression of the density of states at the Fermi energy to 
near zero. The results presented in this report are a natural continuation of such work.

In addition, other systems display such transition. 
Sarma et al.\cite{sarma_cusp_dos} have investigated the electronic 
structure of LaNi$_{1-x}$Mn$_x$O$_3$. It is known that for a critical concentration 
of $x_c \sim 0.1$ a metal-to-insulator transition is found.
The point-contact tunnelling conductance spectra, which provides information on the
density of states, show a  suppression at the Fermi energy in the form of a downward cusp for both 
$x=0.05$ and $0.1$. The conductance is found to be proportional to $\sqrt{V}$, where $V$ is the bias voltage. 
This behaviour was predicted by Altshuler and Aronov\cite{Altshuler_Aronov_gap}, and is similar to
the results that we encountered in studies of LiAl$_y$Ti$_{2-y}$O$_4$.\cite{Fazileh_PRL}
Kim et al.\cite{PhysRevB.71.125104} have studied the transport and optical properties of 
SrTi$_{1-x}$Ru$_x$O$_3$. When $x$ decreases from $1$ to $0$, SrTi$_{1-x}$Ru$_x$O$_3$ evolves from a 
correlated metal, SrRuO$_3$ ($x=1$), to a band insulator, SrTiO$_3$ ($x=0$). Depending on $x$, there are 
six types of electronic states. The metal-to-insulator transition takes place at $x_c \approx 0.5$. 
As $x$ decreases from one, the concentration of conduction electron decreases, and the effective
interaction strength increases. 

These systems are very complicated, having several important $d$ orbitals for different conducting sites,
and many different interaction, disorder and hopping energy parameters.
The theoretical considered below are a minimalist's approach to such interesting behaviour.

\newpage

\subsection{Anderson-Hubbard Model:}

A simplified model that describes interacting electrons moving on a spatially disordered lattice is the 
so-called Anderson-Hubbard model. Its Hamiltonian is given by
\begin{equation}\label{AndersonHubbardHam}
 \hat{\mathcal{H}} = 
 - t \sum_{\langle i j \rangle, \sigma}
 (\hat{c}^{\dag}_{i \sigma} \hat{c}^{}_{j \sigma} + \hat{c}^{\dag}_{j \sigma} \hat{c}^{}_{i \sigma})+ 
\sum_{i,\sigma} \varepsilon_i~\hat{c}^{\dag}_{i \sigma} \hat{c}^{}_{i \sigma}+U \sum_i
 \hat{n}_{i \uparrow} \hat{n}_{i \downarrow}.
\end{equation}
Electron annihilation/creation operators for site $j$ and spin $\sigma=\uparrow,\downarrow$ are represented
by $\hat{c}^{}_{j \sigma}$ and  $\hat{c}^{\dagger}_{j \sigma}$, respectively. The spatially inhomogeneous
environment in which the electrons move is accounted for by
on-site energies, $\varepsilon_i$, which are random. Often, these are selected to be chosen from a uniform distribution,
$\varepsilon_i\in[-\frac W2,\frac W2]$, and therefore $W$ is the energy scale characterizing the strength of the disorder. 
Interactions between electrons are accounted for by the intra-site Hubbard 
interaction term, characterized by the energy scale $U$. The near-neighbour hopping frequency is denoted by 
$t$, and all other symbols have their usual meaning.
It is hoped that this model can capture some of the essential physics of the metal-insulator 
transitions\cite{Imada_MIT} of disordered transition-metal oxides. 

The electronic properties of systems described by this model Hamiltonian are indeed complicated, as can be
understood from the following reasons.
When disorder is sufficiently strong it can lead to the localization
of electrons, and favours large electronic occupancies on sites with low on-site energies. Therefore,
disorder generates both localization effects and an inhomogeneous distribution of electronic charge. There
exists a critical disorder, $(W/t)_c$, beyond which all states are localized and the system becomes an
Anderson insulator.\cite{muC}
In a spatially uniform half-filled system, electron-electron
interactions also lead to the effective localization of electrons;
however, in contrast to disorder, interactions favour single occupancy of electrons on all sites. In general,
at a critical interaction, $(U/t)_c$, the system undergoes a transition from a metal to a Mott insulator.
However, the Mott transition may not take place when the electronic concentration is away from half-filling.

Due to the randomness of the on-site energies, it is usual to address this model using numerical techniques. 
Also, except for some small clusters that can be solved exactly, large lattices have to be solved with 
the help of approximate schemes. This difficulty notwithstanding, a small number of ``exact" numerical 
results are available, coming from both exact diagonalization (ED)\cite{paris:165113, PhysRevLett.83.1826, PhysRevB.63.085102, PhysRevLett.86.2388, epjb.42.279, ImadaPRL} 
and Monte Carlo (MC)\cite{PhysRevB.55.4149, PhysRevLett.83.4610, 
PhysRevB.62.10680, PhysRevB.64.184402, PhysRevLett.87.146401, PhysRevB.67.205112, chakraborty:125117, 
chiesa:086401, paris:165113, Otsuka00} studies. Because of their relevance to our work, we mention one
aspect of the results in two of these papers.
That is, some ED calculations showed
the presence of a suppression of the DOS at the Fermi energy, in both  one\cite{ImadaPRL}
and two\cite{chiesa:086401} dimensions. These results are consistent with the HF papers mentioned below.
Therefore, this provides partial verification of the results based on the HF method, the method
used in the remainder of this paper.

\subsection{Discussion of Previous HF Results:}

The Hartree-Fock (hereafter HF)  
method is well known, and its application to disordered systems is extensive. The accuracy of the
results was recently critiqued by us and coworkers,\cite{chen2008} where it was shown that provided one allowed
for sufficient magnetic degrees of freedom the energies and charge densities of (small) exactly solvable
systems agreed well with those obtained from HF. However, since the spin correlations are essentially
those of pairs of classical spins of variable directions and lengths, and therefore do not include
quantum fluctuations, HF is less successful at capturing the correct spin-spin correlations.

As mentioned, the HF approximation has been applied in many studies of
the Anderson-Hubbard and related models. Some examples are (i) a study of a two-dimensional point-defect model\cite{PhysRevB.47.1126} which represents the acceptors and donors in the high-$T_c$ cuprate La$_{2-x}$Sr$_x$CuO$_4$ through determining the magnetic phase diagram, (ii) a proposal of a novel inhomogeneous metallic phase in two dimensions\cite{Trivedi_PRL_2004, trivedi05a, trivedi05b,kobayashi-2008} as a combined effect of disorder and 
electronic interactions when their strengths are comparable, and (iii) a detailed study of the 
three-dimensional Anderson-Hubbard model, determining both magnetic and electric phase diagrams at 
half-filling.\cite{Tusch_Logan1993, duecker99} 

In Refs. \cite{Tusch_Logan1993,duecker99}, Tusch and Logan have focused on $1/2$-filling and some other fillings lower than $1/8$. At $1/2$-filling with a disorder strength of $W/t=5$, the density of states (hereafter denoted by DOS)
shows a suppression for both $U/t=6$ and $9$. Using the inverse participation ratio (IPR) technique, the system is determined to be metallic at $U/t=6$ and insulating at $U/t=9$. The IPR is compared with a threshold mean IPR (obtained in the noninteracting limit) that scales with system size to determine whether the system is metallic or insulating. However, the effect of the interactions on the threshold mean IPR is unknown, as are the effects
the unusual statistical distribution of the IPR, the latter having been discussed by Mirlin.\cite{mirlin} 
In part to circumvent such problems, here we focus on the characterization of the metallicity of a system
via its optical conductivity.

In the HF treatment of LiAl$_y$Ti$_{2-y}$O$_4$ by one of us and coworkers,\cite{Fazileh_PRL} 
a metal-to-insulator transition, again found from an examination of the behaviour of the IPR, 
was found. In that work the local magnetic moments were restricted to lie along the $z$-axis, 
geometrically frustrated corner-sharing tetrahedral lattices were studied, and 
a quantum site percolation model of disorder was used. Open questions concern whether the suppression 
of the DOS at the Fermi level depends on the lattice type and the disorder model. To answer these questions, 
we choose to study three-dimensional simple cubic lattices, which are unfrustrated, and consider a uniform 
box distribution type of disorder. We also allow local magnetic moments to develop in the $xz$-plane, 
which increases the spin degree of freedom, believed to be important in some circumstances.\cite{chen2008}

We also mention the effective-field theory analysis of local moment formation in disordered
interacting Fermi liquids by Milovanovic, Sachdev, and 
Bhatt.\cite{PhysRevLett.63.82} The potential importance of such moments in the metal-insulator transition 
was left as an outstanding question, and will be one of the main aspects addressed in this paper.

\subsection{Summary of New Results:}

In this paper, we apply the real-space self-consistent HF method to electrons moving on simple cubic 
lattices at $1/4$-filling, with various strengths of interaction and disorder. Here we report our results from
calculations of the DOS and ac conductivity. 

When examining systems with a disorder strength of $W/t=6$, we find that the DOS and the low-frequency 
conductivity are enhanced by a weak interaction ($U/t \le 3$), and for this range of interactions there are 
no magnetic moments in the system, and the ac conductivity is Drude-like. With a stronger interaction 
($U/t \ge 4$), a suppression of the DOS at the Fermi level and qualitative changes (non-Drude-like) 
in the low-frequency conductivity are found. We find that concomitant with these changes in behaviour 
is the appearance of local magnetic moments in the system, although no evidence of magnetic ordering
is present. For this disorder strength and electron concentration
a metal-to-insulator transition is likely to take place at a 
critical $U_c/t \approx 9$; that is, roughly 3/4 of the noninteracting bandwidth. 
We have also examined the weaker disorder strength of $W/t=2$, and although one finds a small suppression
of the DOS with increasing Hubbard interaction, for no value of $U/t$ do we find a metal-to-insulator
transition.

\section{Real-Space Self-Consistent Hartree-Fock Approximation}

For ordered systems, one may transform the Hamiltonian to wave vector space, and then expand 
the many-particle wave function in a complete set of Bloch wave functions in the corresponding 
periodic potential. However, in disordered systems there is no translational symmetry, and working 
in wave vector space does not simplify the problem. Therefore, we will work in real space. In the 
real-space formulation of HF theory ({\em e.g.}, see page 349 of Ref.~\cite{Fazekas}) the 
disorder is treated exactly, and therefore this approximation allows for us to include in our calculations
{\em the effects of localization}. In the real-space formulation of HF theory the local Hubbard 
interaction term is replaced by
\begin{eqnarray}\label{HubbardTerm}
&\hat{n}_{i \uparrow} \hat{n}_{i \downarrow}  =  \hat{c}^{\dag}_{i
\uparrow} \hat{c}^{}_{i \uparrow} \hat{c}^{\dag}_{i \downarrow}
\hat{c}^{}_{i \downarrow}  \nonumber \\
&\approx \langle \hat{n}_{i \downarrow} \rangle \hat{n}_{i \uparrow} +
\langle \hat{n}_{i \uparrow} \rangle \hat{n}_{i \downarrow} -
\langle \hat{n}_{i \downarrow} \rangle \langle \hat{n}_{i \uparrow}
\rangle \\
&- \hat{c}^{\dag}_{i \uparrow} \hat{c}^{}_{i \downarrow} \langle
\hat{c}^{\dag}_{i \downarrow} \hat{c}^{}_{i \uparrow} \rangle -
\hat{c}^{\dag}_{i \downarrow} \hat{c}^{}_{i \uparrow} \langle
\hat{c}^{\dag}_{i \uparrow} \hat{c}^{}_{i \downarrow} \rangle +
\langle \hat{c}^{\dag}_{i \downarrow} \hat{c}^{}_{i \uparrow}
\rangle \langle \hat{c}^{\dag}_{i \uparrow} \hat{c}^{}_{i
\downarrow} \rangle.\nonumber
\end{eqnarray}
Here, the terms that are proportional to fluctuations about the mean values squared are ignored. Also, we 
do not consider the expectation values that arise from superconducting correlations.

Substituting Eq.~(\ref{HubbardTerm}) into the Anderson-Hubbard Hamiltonian in Eq.~(\ref{AndersonHubbardHam}), one finds the HF effective Hamiltonian given by
\begin{eqnarray}\label{effHam}
&\hat{\mathcal{H}}^\textrm{eff} = 
-t\sum_{\langle i j \rangle, \sigma}(\hat{c}^{\dag}_{i \sigma}\hat{c}^{}_{j \sigma}+\hat{c}^{\dag}_{j \sigma} 
\hat{c}^{}_{i \sigma}) \nonumber \\
&+\sum_{i,\sigma} \varepsilon^\textrm{eff}_{i \sigma} \hat{c}^{\dag}_{i \sigma} \hat{c}^{}_{i \sigma}
-U\sum_i(h^{-}_i\hat{c}^{\dag}_{i \uparrow}\hat{c}^{}_{i \downarrow}+h^{+}_i\hat{c}^{\dag}_{i \downarrow}\hat{c}^{}_{i \uparrow}),
\end{eqnarray}
where
\begin{equation}\label{eq:eff1}
\varepsilon^\textrm{eff}_{i \sigma} = \varepsilon_i + U n_{i \bar{\sigma}}~~~~~~~~~
n_{i \sigma} = \langle \hat{n}_{i \sigma} \rangle
\end{equation}
\begin{equation}\label{eq:eff2}
h^{+}_i=\langle \hat{c}^{\dag}_{i \uparrow} \hat{c}^{}_{i \downarrow} \rangle=\langle \hat{S}^+_i \rangle ~~~~~
h^{-}_i=\langle \hat{c}^{\dag}_{i \downarrow} \hat{c}^{}_{i \uparrow} \rangle=\langle \hat{S}^-_i \rangle 
\end{equation}
Here, $n_{i \sigma} = \langle \hat{n}_{i \sigma} \rangle$ is the expectation value of the number of 
electrons with spin $\sigma$ on site $i$; $\varepsilon^\textrm{eff}_{i \sigma}$ is the spin dependent 
effective on-site energy at site $i$, and it is the sum of the original on-site energy plus the 
Hubbard $U$ times the number of electrons with the opposite spin on that site. Therefore, in the absence
of the last term one includes the effects of the Hubbard interaction via an effective shift of the
on-site energies. The last term in Eq.~(\ref{effHam}) does include the $h^{\pm}_i$ effective local fields, and
these guarantee that the effective Hamiltonian is invariant under rotations in spin space.
In the self-consistent formulation of HF theory, one must ensure that the solutions
of the effective one-electron Hamiltonian given in Eq.~(\ref{effHam}) lead to local spin-resolved
charge densities and effective local fields that satisfy Eqs.~(\ref{eq:eff1},\ref{eq:eff2}) when
the expectation values are taken with respect to the HF ground state wave function.

\subsection{Numerical Approach:}

As mentioned above, one is required to solve the effective Hamiltonian self-consistently. In a system with 
$N_s$ sites, one has $4 N_s$ variational parameters that have to be determined numerically so that the 
ground-state energy is minimized. (In addition, one includes a chemical potential to fix the electronic
density at whatever concentration is desired. In our results below we focus on 1/4 filling.)
To solve the effective Hamiltonian self-consistently, one begins with a random initial guess for the 
$4 N_s$ parameters that satisfies the constraint of fixing the total number of electrons, and then
iterates to self consistency. 
One needs to repeat the above procedure with many other initial guesses of the variational parameters
and below we examine the self-consistent states having the lowest energy HF energies, $E_\textrm{HF}$.

After self-consistency is achieved for the HF solutions, we obtain effective single-electron 
energies, \{$E_n$\}, and the DOS is then given by
\begin{equation}\label{dosexp}
    \textrm{DOS}(E) = \sum_n \delta(E - E_n).
\end{equation}
In our results shown below, we broaden each delta function using a Gaussian function given by
\begin{equation}\label{gaussian_func}
   f(x) = \frac{1}{d \sqrt{2 \pi}} e^{-(x - x_0)^2 / 2 d^2},
\end{equation}
where $x_0$ is where $f(x)$ takes maximum value, and $d$ is the standard deviation (or the broadening width).

\begin{widetext}
To calculate the ac conductivity, one first obtains the imaginary part of the current-current correlation function using the Kubo formula:
\begin{equation}\label{imA}
    \textrm{Im} \Lambda_{x x} (\mathbf{q}=0, \omega) 
    = \frac{\pi e^2 t^2}{N_s} \sum_{n, m, n \neq m} \Psi^{n, m} ( f(\epsilon_m) - f(\epsilon_n) ) \delta(\omega - (\epsilon_m - \epsilon_n)),  
\end{equation}
where
\begin{equation}\label{Psinm}
   \Psi^{n,m}=
\sum_{\mathbf{i}, \mathbf{j}, \sigma, \sigma'} (\psi^n_{\mathbf{i} + \hat{x}, \sigma} \psi^m_{\mathbf{i}, \sigma} \psi^m_{\mathbf{j} + \hat{x}, \sigma'} \psi^n_{\mathbf{j}, \sigma'} - \psi^n_{\mathbf{i} + \hat{x}, \sigma} \psi^m_{\mathbf{i}, \sigma} \psi^m_{\mathbf{j}, \sigma'} \psi^n_{\mathbf{j} + \hat{x}, \sigma'}
-\psi^n_{\mathbf{i}, \sigma} \psi^m_{\mathbf{i} + \hat{x}, \sigma} \psi^m_{\mathbf{j} + \hat{x}, \sigma'} \psi^n_{\mathbf{j}, \sigma'} + \psi^n_{\mathbf{i}, \sigma} \psi^m_{\mathbf{i} + \hat{x}, \sigma} \psi^m_{\mathbf{j}, \sigma'} \psi^n_{\mathbf{j} + \hat{x}, \sigma'}).
\end{equation}
Then, the real part of the ac conductivity is\cite{Mahanbook}
\begin{equation}\label{acconductivity}
    \sigma_1 (\omega) = \frac{\text{Im} \Lambda_{x x} (\mathbf{q}=0, \omega)}{\omega}.
\end{equation}
To calculate the ac conductivity numerically, we broaden each delta function in Eq.~(\ref{imA}) using a Gaussian function given by Eq.~(\ref{gaussian_func}). In our results below we state the broadening for each quantity.
\end{widetext}

\section{Results}

\subsection{Disorder Strength:}

While the Anderson-Hubbard model is often used to study disordered electronic systems with strong
electron-electron interactions, it is not always clear how the parameter space studied for such models 
relates to real materials, particularly with regards to the strength of the disorder potential. 
Here we focus on the appropriate range of disorder strengths.

When characterizing the strength of the potential energy it is usual to compare $U$ to the noninteracting
bandwidth for electrons moving on an ordered lattice. For a 3d simple cubic lattice that is $B=12t$. Weak
interactions correspond to $U/B \ll 1$, intermediate to $U/B \sim 1$, and strong to  $U/B \gg 1$. This
leads to the question, is it permissible to use the same characterization for the disorder strength?

We have examined the $U=0$ Anderson model (at 1/4 filling - the same as the electronic concentration 
studied throughout this paper) using a box distribution disorder -- all on-site energies between
$-W/2$ and $+W/2$ are equally probable. We have then found the ac conductivity of this system for
various $W/t$, and here we discuss the results corresponding to $W/t=1, 2$ and 6. 
If one fits our conductivity data to the real conductivity of a Drude model, the latter given by
\begin{equation}\label{eq:Drude-real}
\sigma_1(\omega)~=\frac{\sigma_{dc}}{1+(\omega\tau)^2}~~~,
\end{equation}
one obtains estimates of the relaxation time and the dc conductivity shown in Table~\ref{table:Drude}.
The difference between
the $\omega\rightarrow 0$ extrapolations quantifying the dc conductivities are striking,
in that for $W/t=1$ we find $\sigma_{dc} \approx 15 \pi e^2 t^2$, whereas for $W/t=6$ we find 
$\sigma_{dc} \approx 0.26 \pi e^2 t^2$, implying that the conductivity of these systems differs by
almost a factor of 60. Clearly, the noninteracting $W/t=6$ (or equivalently $W/B=1/2$) system corresponds 
to a ``bad metal". (However, while the $W/t=6$ system is a very poor conductor, for 
this disorder strength we are not approaching a 1/4-filled Anderson insulator,
and, in fact, our calculations and existing data\cite{muC} allow us to determine 
that $\varepsilon_F-\mu_C \sim 4t$, with $\mu_C$ being the location of the mobility edge.)

\begin{table}
\caption{\label{table:Drude} 
Conducting properties of noninteracting electrons moving on a 3d simple cubic lattice
with a box distribution of disorder of strength $W$, for an electron filling of 1/4. The
$\sigma_{dc}$ and $\tau$ are found from fits of the ac conductivity to Eq.~(\protect\ref{eq:Drude-real}).}
 \begin{ruledtabular}
  \begin{tabular}{ccc}
$W/t$ & $\sigma_{dc}$ [$\pi e^2 t^2$] & $\tau$ [$t^{-1}$] \\
\hline
1 & 15. & 11.  \\
2 & 2.3 & 3.6  \\
6 & 0.26 & 0.45 \\
  \end{tabular}
 \end{ruledtabular}
\end{table}

Further justification for this claims follows from a
second way of characterizing the ``strength" of the disorder, and corresponds to evaluating the
elastic mean free path, $\ell_e$, in the Born approximation. One finds\cite{Mott_MIT}
\begin{equation}\label{eq:mfp}
\frac{\ell_e}{a}~\approx~\frac{4\pi}{z^2}~\Big(\frac BW\Big)^2~~~,
\end{equation}
where $a$ is the lattice constant and $z$ is the coordination number for a particular lattice. 
Therefore, for $W/t=1$,~2, and 6 one has ${\ell_e}/a\approx 50$, 13, and 1.4, respectively. 
While one does not expect the Born approximation to be accurate quantitatively, especially
when it predicts  ${\ell_e}\approx a$, from these results it follows that
for noninteracting electrons moving on a three dimensional simple cubic lattice 
with $W/B=1/2$ one has a very short mean free path.
So, while it might be conventional to refer to a $U/B=1/2$ correlated electronic
systems as being intermediate coupling, for a disorder strength of $W/B=1/2$ the system remains metallic
but the effect of the disorder potential is indeed very large. As we see in the results below, it is only
for such large disorder strengths, which by itself leads to substantial localization effects,
that we find a disorder+correlation-driven metal-insulator transition. 

(Also, note that in our previous study of lithium titanate\cite{Fazileh_PRL} 
we employed a quantum site percolation model, which formally
corresponds to setting all A site energies to be zero while those on the B sites are infinite - that
is, the B sites are removed from the conduction path. With such a large disorder potential 
(of infinite strength for a binary alloy model of disorder) we
did obtain a metal-insulator transition.\cite{Fazileh_PRL})

\subsection{Variation of the Density of States:}

As mentioned in the introduction, one of the motivations for this study was to better understand the conditions
necessary for the appearance of a suppression in the DOS at the Fermi level. Previous
work\cite{Fazileh_PRL, Farhad_Thesis} was performed on a lattice appropriate for the 
description of LiAl$y$Ti$_{2-y}$O$_4$, namely
on a corner-sharing tetrahedral lattice. Also, that work was completed using a restricted HF theory: 
In terms of the above-presented formalism, the effective local fields $h^\pm_i$ were set to be zero, meaning
that the spin degrees of freedom were restricted to develop along the quantization axis ($z$). 

All of our results here are for a 3d simple cubic lattice -- unlike the corner-sharing tetrahedral lattice,
this lattice is bipartite and therefore unfrustrated. In terms of the magnetic degrees of freedom, here we 
present results from our new HF calculations for two different situations. 
First, we further restrict the system to be paramagnetic, meaning that we do not allow for the formation of
{\em any} local moments -- this means that $n_{i\uparrow}$ is forced to be equal to $n_{i\downarrow}$. 
Second, we relax the restriction of the moments being along the $z$-axis, and allow them to form in the 
$xz$-plane, i.e., $h_i^\pm$ is nonzero but real.

\begin{figure}[ht]
  \begin{center}
    \includegraphics[width=7.5 cm, clip=true]{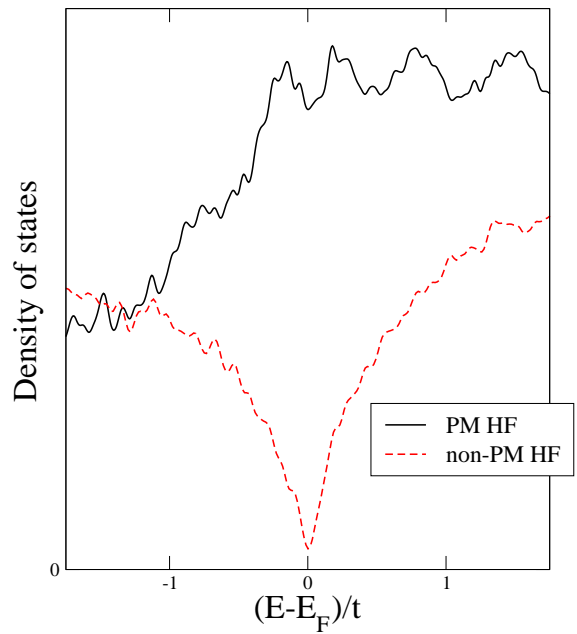}\\
    \caption{[Colour online] Density of states of a three-dimensional simple cubic lattice of size $18^3$ for an
electronic filling of $1/4$. The interaction strength is $U/t=11.5$, and the disorder strength is 
$W/t=6$. The nonparamagnetic HF result (dashed curve) shows a strong suppression of the DOS at 
the Fermi level, whereas the paramagnetic HF result (solid curve) shows no such suppression. The
broadening used corresponds to $d=0.024t$.}
    \label{DOS_3dSC_L18_Vdis6.0_U11.5_edensity0.5}
  \end{center}
\end{figure}

Some of our results for the DOS for a three-dimensional simple cubic lattice 
are shown in Fig.~\ref{DOS_3dSC_L18_Vdis6.0_U11.5_edensity0.5}. The lattice size is $18^3$, 
and the electronic filling factor is $1/4$. The interaction strength is $U/t=11.5$, where the DOS
at the Fermi level has its largest suppression, and the disorder is modelled with a box distribution
for a disorder strength of $W/t=6$. 
The nonparamagnetic HF result shows a strong suppression of the DOS at the Fermi level; 
however, the paramagnetic HF result shows no suppression at all. 

Since we only found a suppression of the DOS with nonparamagnetic HF solutions, both with
$h_i^\pm = 0$ (in Ref.~\cite{Fazileh_PRL}) and $h_i^\pm$ being real, 
the magnetic moments are essential to the suppression and to the 
metal-insulator transition (at least within the HF context). The moments found for the real $h_i^\pm$ HF 
ground state are strongly noncollinear -- \textit{e.g.}, see our discussion in Ref.~\cite{chen2008}.
Therefore, the restriction of collinear moments employed in 
Refs.~\cite{Fazileh_PRL} is not important to the suppression. Also, the type of lattice 
does not matter because the suppression is found both for the frustrated corner-sharing tetrahedral 
lattices (in Ref.~\cite{Fazileh_PRL}) and for the unfrustrated simple cubic lattices 
studied in this paper. Also, the type of disorder does not matter, since the suppression is found either 
with a quantum site percolation model (in Ref.~\cite{Fazileh_PRL}) or with a uniform 
box distribution type of disorder.

To better quantify the presence of the local moments we have calculated the average magnitude of the 
moment per electron. We define an Edwards-Anderson-like order parameter
\begin{equation}\label{qEA}
   {\overline m} = \frac{1}{N_e} \sum_{i=1}^{N_s} |\langle\hat{\mathbf{S}}_i\rangle|,~~~~~
\hat{\mathbf{S}}_i\equiv\frac12 
\hat{c}^\dagger_{i\sigma}{\vec \tau}_{\sigma\sigma^\prime}\hat{c}_{i\sigma^\prime}
\end{equation}
where $\hat{\mathbf{S}}_i$ is the spin operator on site $i$, ${\vec \tau}$ are the Pauli matrices,
and $N_e$ ($N_s$) is the number of electrons (lattice sites).
The quantity ${\overline m}$ is similar to the Edwards-Anderson order parameter\cite{EA} 
in spin glass theory, which is used to distinguish between glass and nonglass
phases,\cite{PhysRevB.38.9093} but here we use ${\overline m}$ as a characterization of local magnetic 
moment formation. 

\begin{figure}[ht]
  \begin{center}
    \includegraphics[width=8 cm, clip=true]{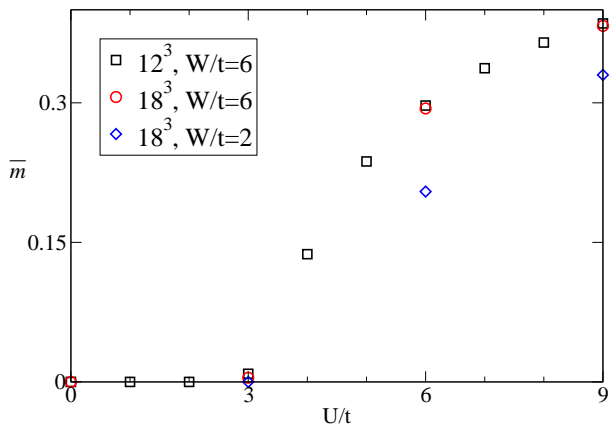}\\
    \caption{[Colour online] 
The average magnitude of the magnetic moment per electron (${\overline m}$) {\em vs.} 
interaction strength ($U/t$) for a $12^3$ lattice with disorder strength $W/t=6$ (squares), 
and for an $18^3$ lattice with $W/t=2$ (diamonds) and $6$ (circles).}\label{LocalMoments}
  \end{center}
\end{figure}

We have considered many different parameter sets and lattice sites (see discussions below), and our
results for ${\overline m}$ are shown in Fig.~\ref{LocalMoments}. We see
that are no local moments in the noninteracting ($U/t=0$) or weakly interacting systems ($U/t \le 3$) 
for $W/t~=~2$ and 6. For the $18^3$ lattice and $U/t=3$ and $W/t=6$, ${\overline m}$ is 
only about $0.0045$, and for the $18^3$ lattice and $U/t=3$ and $W/t=2$ we find 
${\overline m} = 0.00002$; therefore, this quantity is expected to be zero in the thermodynamic limit. 
For larger $U/t$ we do find moments, and for a given interaction strength ${\overline m}$ 
increases as one increases the strength of the disorder. In all of the
subsequent data that we show having a suppression of the DOS at the Fermi level, and non-Drude-like
conductivity, the average magnitude of the local magnetic moments $\overline m$ \textbf{is always nonzero.}

\begin{figure}[ht]
  \begin{center}
    \includegraphics[width=8 cm, clip=true]{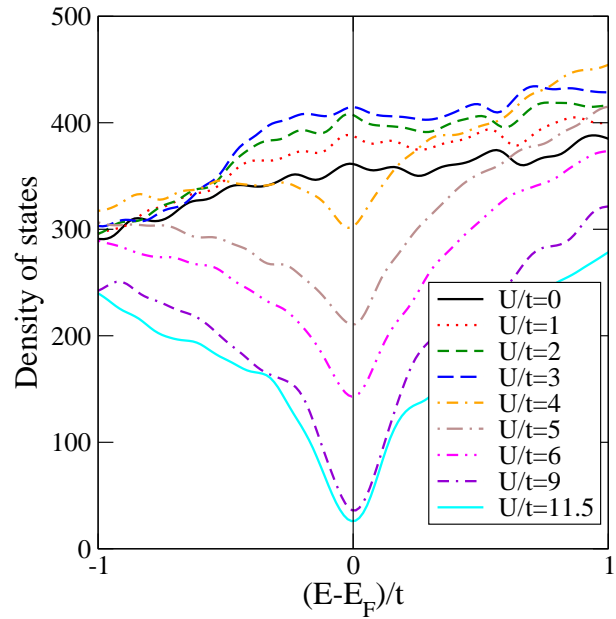}\\
    \caption{[Colour online]
Density of states of the Anderson-Hubbard model on a $1/4$-filled three-dimensional simple 
cubic lattice of size $12^3$ for a disorder strength of $W/t=6$. Results without interactions ($U/t=0$) 
and with interaction strengths $U/t=1$, $2$, \ldots, $6$, $9$, and $11.5$ are averaged over $4$ 
realizations of spatial disorder. The vertical line represents the Fermi level. The broadening width 
is $d = 0.048 t$.}\label{DOS_L12_W6_box}
  \end{center}
\end{figure}

Now, we consider the variation of the DOS in more detail. We have calculated the DOS 
for a $1/4$-filled three-dimensional simple cubic lattice of size $12^3$ for various strengths of interaction 
$U/t=0$, $1$, $2$, \ldots, $8$, $9$, and $11.5$. The on-site energies obey a uniform box distribution, and 
the disorder strength is $W/t=6$. Results are averaged over $4$ realizations of disorder -- a limited study
using more realizations of disorder found no qualitative changes when more realizations were employed. 
The maximum differences for the self-consistency are between $2 \times 10^{-5}$ (for $U/t=1$ and $2$) 
and $1.2 \times 10^{-3}$ (for $U/t=11.5$). (Typically, the average differences were at least one order
magnitude smaller than the maximum differences.) In Fig.~\ref{DOS_L12_W6_box}, we show the DOS 
of this system obtained with $d=0.048 t$. (The DOS curves are smooth and well separated from 
one another around the Fermi level, whereas for smaller broadening these curves are bumpy and not well 
separated, and we therefore choose $d = 0.048 t$ as the broadening width for the DOS of this system.)

As shown in Fig.~\ref{DOS_L12_W6_box}, for $U/t=0$ to $3$, each DOS does not show any suppression. In fact, the DOS at the Fermi level is enhanced as the interaction is turned on and then increased. 
The suppression appears for the first time when $U/t = 4$, and the amount of the suppression increases 
as $U/t$ is increased. We find that the maximum suppression, for 1/4 filling, occurs around $U/t=11.5$,
although there is little difference between the DOS for this $U/t$ and other Hubbard interactions 
close to this strength.

\subsection{Behaviour of the Optical Conductivity:}

Extracting the low-frequency behaviour of the conductivity is nontrivial, and we briefly outline 
the numerical approach taken. We have examined the ac conductivity obtained with different values 
of the Gaussian broadening width $d$ in Eq.~(\ref{gaussian_func}), and have selected a value for $d$
based on the following: For a three-dimensional simple cubic lattice of size $12^3$ with disorder 
strength $W/t=6$ and interaction strength $U/t=6$, we calculated the ac conductivity with broadening 
widths $d = 0.012t$, $0.03t$, and $0.06t$ (results not shown). Because the broadening width $d$ has 
a finite value, the contributions of frequencies between $0$ and $d$ makes the imaginary part of the 
current-current correlation function 
$\textrm{Im} \Lambda(\mathbf{q}=0, \omega) = \omega \sigma_1(\omega)$ nonzero around $\omega = 0$. 
As a result, when calculating $\sigma_1(\omega)$ by dividing $\textrm{Im} \Lambda(\mathbf{q}=0, \omega)$ 
by $\omega$, we obtain diverging $\sigma_1(\omega)$ around $\omega = 0$. The divergence is simply
an artifact of numerical procedure used (a broadening width $d$ that is finite), and is not associated 
with the physics of the system being investigated. Therefore, we have to cut off the conductivity curves 
at low frequencies ($\sim d$) where they ``turn up''. Recalling that the DOS data of 
Fig.~\ref{DOS_L12_W6_box} used $d=0.048t$, we choose $d = 0.03t$ as the broadening width for
the conductivity because the corresponding conductivity curves are smooth and still retain much of 
the low-frequency behaviour -- simply, we are not forced to discard as much low-frequency data as we would 
with $d=0.048t$. (As reviewed in the discussion, this low-frequency behaviour is an important quantity 
to know.)

We now discuss our results, juxtaposing DOS and ac conductivity data for each parameter set.
First, we discuss results for studies done on lattices of size $12^3$ with a disorder strength of $W/t = 6$; 
an average over $4$ realizations of disorder is used for each data set.
We increase the interaction strength $U/t$ from zero up to roughly $12$.

\begin{figure}[h]
  \begin{center}
    \includegraphics[width=8 cm, clip=true]{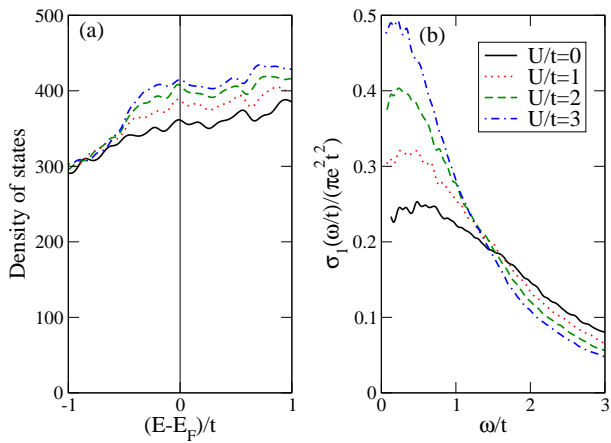}\\
    \caption{[Colour online] Density of states (a) and ac conductivity (b) for a $1/4$-filled three-dimensional 
simple cubic lattice of size $12^3$ with a disorder strength $W/t = 6$  with no interactions 
($U/t=0$) and with Hubbard interaction of strengths $U/t=1$, $2$, and $3$. The DOS do not show 
any suppression at the Fermi level, and in fact increase as the interaction strength $U/t$ increases.
The low-frequency conductivity is enhanced as $U/t$ is increased, and the ac conductivity is Drude-like.
We have used a broadening corresponding to $d=0.048t$ for the DOS and $d=0.03t$ for the conductivity.}
    \label{dos_conductivity_L12_Ne864_W6_U0_1_2_3}
  \end{center}
\end{figure}

The results for the noninteracting case ($U/t=0$) and for interactions $U/t = 1$, $2$, and $3$ are shown in Fig.~\ref{dos_conductivity_L12_Ne864_W6_U0_1_2_3}. The solid lines in (a) and (b) represent the DOS and the 
ac conductivity for the noninteracting electrons, respectively. The ac conductivity has a shape that is 
typical of a metal, but the low-frequency peak is broad.  When we turn on the interaction to $U/t = 1$ 
(dotted lines), we see an enhancement in the DOS at the Fermi level. Concomitantly, the low-frequency 
ac conductivity also increases from the noninteracting value.
These enhancements could be a result of the screening of the disorder by the Hubbard interactions. 
As the interaction strength $U/t$ increases to $2$ (dashed lines) and $3$ (dash-dotted lines), 
we find that the DOS at the Fermi level and the low-frequency ac conductivity are 
both further enhanced. Therefore, for a disorder of $W/t = 6$ at 1/4 filling, a weak interaction 
of $U/t \le 3$ enhances the low-frequency ac conductivity due to an increase in the DOS at the Fermi level.
Recall, from Fig.~\ref{LocalMoments}, that for this range of interaction strengths no
local moments are formed.

\begin{figure}[h]
\begin{center}
\includegraphics[width=8 cm, clip=true]{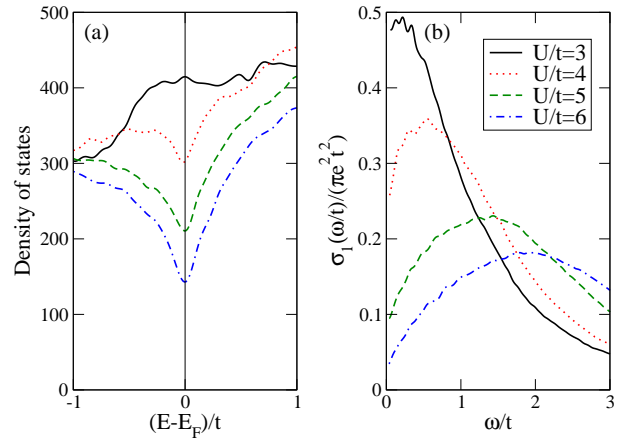}\\
\caption{[Colour online] Density of states (a) and ac conductivity (b) for a $1/4$-filled three-dimensional 
simple cubic lattice of size $12^3$ with a disorder strength $W/t = 6$ and Hubbard interaction strengths
$U/t=3$, $4$, $5$, and $6$. Each of the DOS for $U/t=4$, $5$, and $6$ shows a suppression at the
Fermi level, and the amount of the suppression increases as $U/t$ increases.
The low-frequency conductivity decreases as $U/t$ increases, and the ac conductivity is no longer Drude-like.}
    \label{dos_conductivity_L12_Ne864_W6_U3_4_5_6}
  \end{center}
\end{figure}

The results for interactions $U/t=3$, $4$, $5$, and $6$ are shown in Fig.~\ref{dos_conductivity_L12_Ne864_W6_U3_4_5_6}. A suppression of the DOS at the Fermi level first appears for $U/t=4$, and its value is smaller than that of the noninteracting electrons and that of the $U/t=3$ system. The low-frequency conductivity for $U/t = 4$ is also smaller than that of the $U/t = 3$ system. As the interaction strength $U/t$ increases to the values of $5$ and $6$, the amount of the suppression of the DOS at the Fermi level increases, and the low-frequency conductivity decreases. Each of the ac conductivity curves still extrapolates to a nonzero value when the frequency $\omega / t \rightarrow 0$. We note that starting from $U/t = 5$, the low-frequency conductivity is smaller than that of the noninteracting electrons. More importantly, the ac conductivity for $U/t=4$, $5$, and $6$ is no longer Drude-like.
In fact, our conductivity curves are similar to the localization-enhanced Drude theory of Mott and 
Kaveh,\cite{MottKaveh} and such conductivities have been observed experimentally previously in the metallic 
phase near the metal-insulator transition in experimental work on conducting polymers.\cite{PhysRevB.48.14884}

\begin{figure}[ht]
  \begin{center}
    \includegraphics[width=8 cm, clip=true]{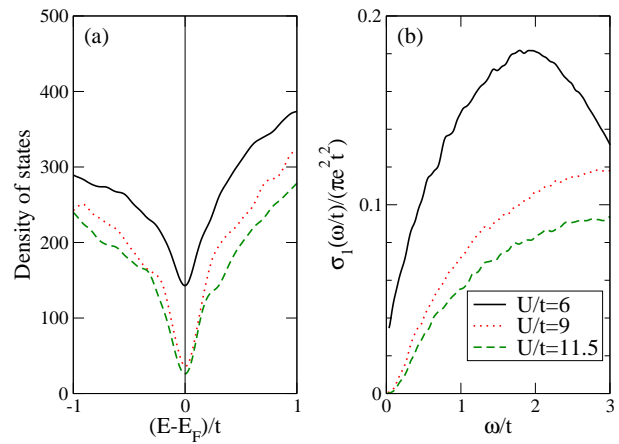}\\
    \caption{[Colour online] 
Density of states (a) and ac conductivity (b) for a $1/4$-filled three-dimensional simple 
cubic lattice of size $12^3$ with a disorder strength $W/t = 6$ and Hubbard interactions of strength 
$U/t=6$, $9$, and $11.5$. Each DOS shows a suppression at the Fermi level, and the low-frequency 
conductivity is also suppressed -- both suppressions increase with increasing interactions. 
The ac conductivity for $U/t=6$ extrapolates to a nonzero dc conductivity as $\omega/t \to 0$, however, the dc conductivity for $U/t=9$ and $11.5$ are both very close to zero.}
    \label{dos_conductivity_L12_Ne864_W6_U6_9_11.5}
  \end{center}
\end{figure}

We show the results for interaction strengths $U/t=6$, $9$, and $11.5$ in Fig.~\ref{dos_conductivity_L12_Ne864_W6_U6_9_11.5}. Compared to $U/t=6$, the DOS at the Fermi level gets further suppressed. However, the amount of the suppression for $U/t=9$ and $11.5$ are not that different. It is very clear that, on the scale of Fig.~\ref{dos_conductivity_L12_Ne864_W6_U6_9_11.5}, the dc conductivity for $U/t=9$ and $11.5$ are both zero. The ac conductivity for $U/t=9$ and $11.5$ both increase as $\omega / t$ increases up to the value of $\omega / t \sim 3$.

We fit the ac conductivity for $U/t=6$, $9$, and $11.5$ with power-law relations. For $U/t=6$, the system appears to be metallic, and we fit its ac conductivity using the following equation:
\begin{equation}\label{metallic_powerlaw}
    \sigma_1 (\omega) = \pi e^2 t^2  (\sigma_0 + \beta \omega^\alpha),
\end{equation}
where $\sigma_0$, $\beta$, and $\alpha$ are the parameters to be determined. The range of frequency 
over which we chose to fit the data is $0.04 < \omega/t < 1.2$ (recall that we used $d=0.03t$ in producing
these conductivity curves), and we obtain 
\begin{eqnarray}\label{metallic_powerlaw_fit}
   & \sigma_1 (\omega/t) / (\pi e^2 t^2) = (0.002 \pm 0.001) \nonumber \\
   & + (0.145 \pm 0.001) (\omega/t)^{0.501 \pm 0.007}.
\end{eqnarray}
Here, the uncertainties are one standard deviation. We learn from this power-law fit that the dc conductivity for the system with $W/t = U/t = 6$ is $\sigma_0 = 0.002 \pm 0.001$, which is quite close to but still above zero. Therefore, we expect that the system is a metal. We also see that the exponent of the frequency is $\alpha = 0.501 \pm 0.007$, 
which is the exponent ($0.5$) that appears in the ac conductivity of noninteracting electrons with 
strong disorder, in the metallic regime close to the transition,\cite{Kroha1990231} 
where $\sigma_1 (\omega)$ near $\omega = 0$ goes as $\sigma_1 (\omega) - \sigma_0 \sim \sqrt{\omega}$.

We fit the ac conductivity for $U/t=9$ and $11.5$ 
over the ranges of $\omega/t \in [0.074, 0.267]$ and $\omega/t \in [0.074, 0.330]$, respectively. Setting
$\sigma_0=0$ we obtain
\begin{equation}\label{insulating_powerlaw_fit_U9}
    \sigma_1 (\omega/t) / (\pi e^2 t^2) = (0.23 \pm 0.01) (\omega/t)^{1.89 \pm 0.02}
\end{equation}
for $U/t=9$
and
\begin{equation}\label{insulating_powerlaw_fit_U11.5}
    \sigma_1 (\omega/t) / (\pi e^2 t^2) = (0.178 \pm 0.006) (\omega/t)^{2.05 \pm 0.02}
\end{equation}
for $U/t=11.5$.
These exponents are very close to $2$, and, again, this is the same as the exponent for noninteracting electrons 
for an even stronger disorder, namely for a disordered system in the insulating phase,\cite{Kroha1990231} 
where $\sigma_1 (\omega) \sim \omega^{2}$ in the low-frequency regime.

Therefore, the system with $W/t=6$ is metallic for $U/t \le 6$ and insulating for $U \gtrsim 9$. By examining
other $U/t$ around 9 (data not shown) we can identify that the metal-to-insulator transition for this disorder 
and electronic filling indeed takes place at some critical $(U/t)_{\textrm{cr}}\approx 9$.

It is always desirable to complete studies with larger lattices, where finite-size effects are 
hopefully less punitive. In our study the benefits of having such data are (i) better energy resolution of the
DOS near the Fermi level, and below we make clear the usefulness of such data;
and (ii) better resolution of the ac conductivity at low frequencies.
Using our computer resources the largest lattice that we have managed to treat, and are able
to explore various parameters sets, is an $18^3$ lattice with $1/4$-filling. For this lattice with
our HF formulation employing real $h_i$, 
it takes about \textit{one month} to achieve self-consistency in the HF calculations 
for a single realization of disorder, and due to this enormous amount of time we solve the problem 
only for one realization of disorder. 

\begin{figure}[ht]
  \begin{center}
    \includegraphics[width=8 cm, clip=true]{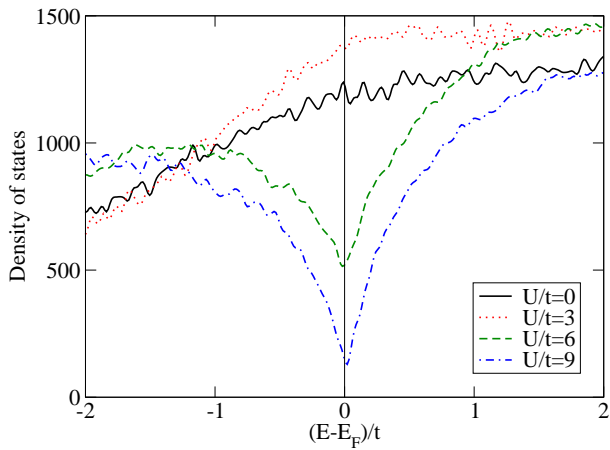}\\
    \caption{[Colour online]
Density of states for the Anderson-Hubbard model on a $1/4$-filled three-dimensional simple cubic lattice of size $18^3$. For an intermediate disorder of $W/t=6$, interaction strengths of $U/t=0$, $3$, $6$, and $9$ are considered. The Gaussian broadening width is $d=0.024 t$.}\label{DOS_L18_W6_box}
  \end{center}
\end{figure}

For a disorder strength of $W/t=6$ and interaction strengths of $U/t=0$, $3$, $6$, and $9$,
we show the resulting DOS in Fig.~\ref{DOS_L18_W6_box}. (Due to the higher number of energy
levels per unit frequency we have used 1/2 the broadening width ($d=0.024t$) as we did for 
the 12$^3$ studies.) We note that the behaviour of the DOS in this larger lattice is very 
similar to the $12^3$ lattice that we have studied earlier. 

\begin{figure}[h]
  \begin{center}
    \includegraphics[width=8 cm, clip=true]{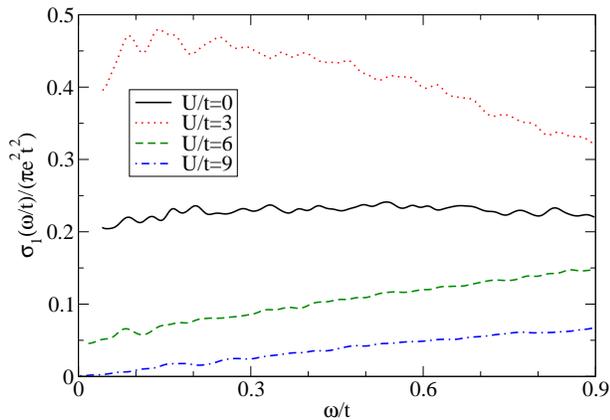}\\
    \caption{[Colour online]
AC conductivity of the Anderson-Hubbard model on a $1/4$-filled three-dimensional simple 
cubic lattice of size $18^3$ for a disorder potential of strength $W/t=6$, for interaction strengths of
$U/t=0$, $3$, $6$, and $9$. The Gaussian broadening width is $d=0.01 t$.}\label{conductivity_L18_W6_box}
  \end{center}
\end{figure}

The corresponding ac conductivity for the same system is shown in Fig.~\ref{conductivity_L18_W6_box}.
Note that this data corresponds to a very small Gaussian broadening of only $d=0.01 t$, and
therefore we can obtain data down to very low frequencies. We clearly see that the data is similar (qualitatively
{\em and} quantitatively, indicating that our data does not suffer from strong finite-size effects) to
that shown earlier for a 12$^3$ lattice. Further, the extrapolation of the conductivity in the zero frequency 
limit makes clear the nonconducting nature of the $U/t=9$ system.

\subsection{System with a weak disorder:}

So far, we have studied systems with a disorder strength of $W/t=6$. As mentioned earlier, this corresponds to
a noninteracting disordered electronic system with a very low dc conductivity -- that is, this corresponds to
a ``bad metal".  We found a suppression of the DOS at the Fermi level for interaction strengths $U/t \ge 4$.
When this happens, we find that the ac conductivity is no longer Drude-like, and for strong enough interactions, 
$U/t\sim 9$, the system no longer possesses a metallic conductivity. A natural question then
arises: For smaller strengths of disorder does one still obtain a metal-insulator transition?

In this subsection we present our results for a weaker disorder corresponding to $W/t=2$. 
Compared to the noninteracting $W/t=6$ system, the noninteracting $W/t=2$ system has a mean free path 
and dc conductivity almost an order of magnitude larger.

\begin{figure}[ht]
  \begin{center}
    \includegraphics[width=8cm, clip=true]{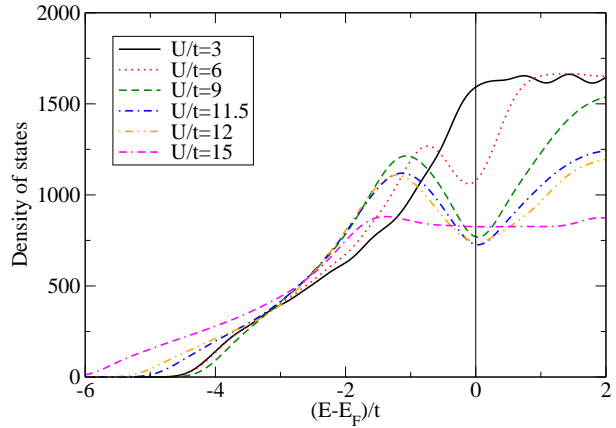}\\
    \caption{[Colour online]
Density of states for a $1/4$-filled three-dimensional simple cubic lattice of size $18^3$ with a box distribution type of disorder of strength $W/t = 2$ ($4$ realizations of disorder) and Hubbard interaction of strengths $U/t=3$, $6$, $9$, $11.5$, $12$, and $15$.}\label{dos_L18_W2}
  \end{center}
\end{figure}

For such disorder we are forced
to use a large lattice because the energies are strongly degenerate in small lattices. 
In Fig.~\ref{dos_L18_W2} we show the DOS for an $18^3$ simple cubic lattice with interaction strengths 
$U/t=3$, $6$, $9$, $11.5$, $12$, and $15$. 

For this weak disorder strength, the interaction strength of $U/t=3$ does not produce a suppression in the DOS at the Fermi level. However, for $U/t=6$, $9$, $11.5$ and $12$ a suppression in the DOS appears at the Fermi level, and the amount of suppression is deepest at $U/t=11.5$. For $U/t=15$ we find that the suppression has been
eliminated.

\begin{figure}[h]
  \begin{center}
    \includegraphics[width=8cm, clip=true]{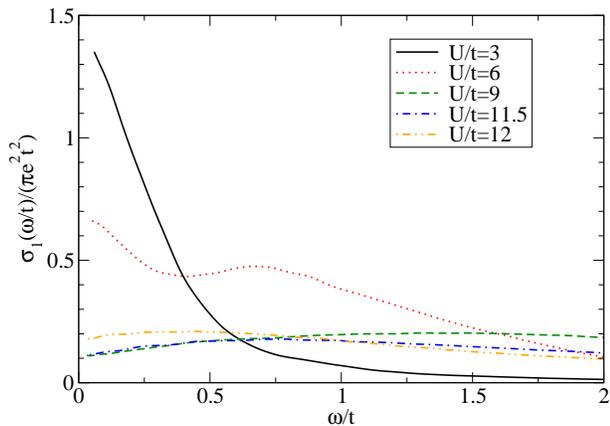}\\
    \caption{[Colour online]
AC conductivity for a $1/4$-filled three-dimensional simple cubic lattice of size $18^3$ with a box distribution type of disorder of strength $W/t = 2$ ($4$ realizations of disorder) and Hubbard interaction of strengths $U/t=3$, $6$, $9$, $11.5$, and $12$.}\label{conductivity_L18_W2}
  \end{center}
\end{figure}

In Fig.~\ref{conductivity_L18_W2}, we show the ac conductivity for the same system with interaction 
strengths up to $U/t=12$. The conductivity for $U/t=3$ is clearly metallic and Drude like,
and extrapolates to roughly $\sigma_{dc} \approx 1.4 \pi e^2 t^2$ in the low-frequency limit. This value is 
roughly half that of noninteracting electrons with the same strength of disorder.

With increasing $U/t$ the conductivity at low frequencies is lowered, 
with $U/t=9$ and 11.5 having very similar behaviour. 
However, for $U/t=12$ the conductivity increases near $\omega=0$ -- this behaviour persists for even 
larger Hubbard energies. Therefore, this disorder strength 
does not lead to a nonmetallic conductivity for any Hubbard energy that we have studied.

\begin{figure}[h]
  \centering
  \includegraphics[width=8cm, clip=true]{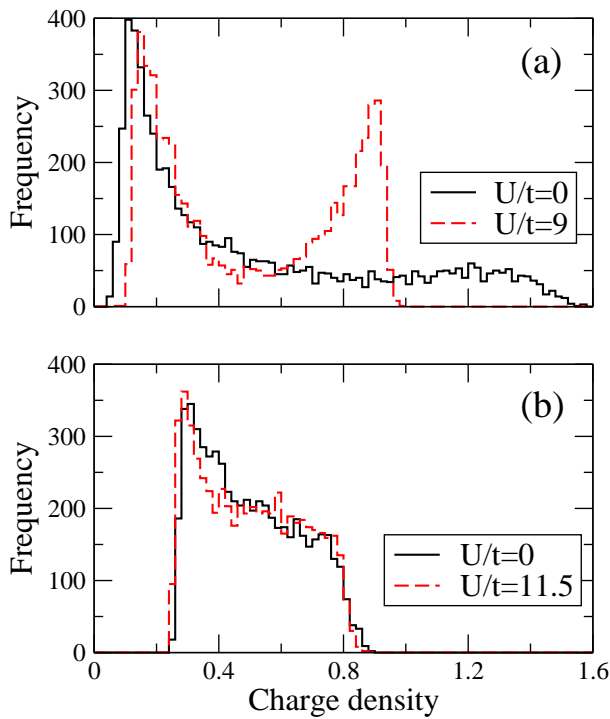}\\
  \caption{[Colour online]
Histograms of the local charge density for the $18^3$ lattice. (a) $U/t=0$ and $9$ for a 
disorder $W/t=6$. (b) $U/t=0$ and $11.5$ for a weak disorder $W/t=2$.}\label{charge_hist_L18_W2and6}
\end{figure}

To aid in the understanding the above results, namely that for a larger disorder one does obtain a metal-insulator
transition, whereas for smaller disorder one does not, we have examined the local charge densities for these
parameter sets. In Fig.~\ref{charge_hist_L18_W2and6}, we show histograms of the local charge density 
for the $18^3$ lattice with disorder strength $W/t=6$ for $U/t=0$ and $9$ (a), and $W/t=2$ for $U/t=0$ and 
$11.5$ (b).

For disorder strength $W/t=6$, when the system is noninteracting the charge density is spread out over a large
range, with one large peak close to $n\sim0.1$. That is, in the absence of interactions and for this seemingly
large disorder (see previous discussion) the ground state corresponds to a very small occupation of sites 
with large on-site energies, with a increasing occupation as the on-site energies decrease. 
However, when the interaction strength is strong ($U/t=9$), the original peak at the small charge density 
does not change much, but a large new peak around $n \approx 0.9$ forms. That is, for this disorder
the interactions lead to very different charge densities, and correspond to those of an insulating system.
The contrasting situation for weak disorder is clear from the figure --  the charge distribution is quite 
similar between the noninteracting ($U/t=0$) and strongly interacting ($U/t=11.5$) systems, and in
both cases one has metallic conduction.

We have also plotted the histograms for the effective on-site energies, that is 
$\varepsilon^\textrm{eff}_{i \sigma}$ from Eq.~(\ref{eq:eff1}) averaged over both spins, since this
quantity has been suggested to be important in the understanding of the metallicity of such
ground states.\cite{Bill2008,Kroha2008} However, the histograms for 
$\langle\varepsilon^\textrm{eff}_{i \sigma}\rangle_\sigma$, 
for the same parameter sets as those used in Fig.~\ref{charge_hist_L18_W2and6}, show essentially 
no difference for these interacting HF ground states, one of which we found to be metallic and one insulating.

\section{Discussion}\label{sec:discussion}

Our numerical analysis of the real-space self-consistent HF treatment of
the Anderson-Hubbard model away from 1/2 filling on 3d simple cubic lattices gives rise to
three main conclusions, each of which we discuss in relation to previously published theoretical
and experimental work.

\begin{itemize}
\item At least at 1/4 filling, and thereby well away from 1/2 filling, one requires sufficiently 
strong disorder to obtain a metal-to-insulator transition.
\end{itemize}

As preliminary work to our study, we first characterized the consequences of different strengths 
of disorder for noninteracting electrons. For a disorder strength much less than the 
noninteracting bandwidth, given by $B=12t$, we
find a large dc conductivity, consistent with theoretical predictions of a large mean free path. However.
for a disorder strength that is 1/2 of this bandwidth, we find a greatly reduced dc conductivity,
and applying Eq.~(\ref{eq:mfp}), estimate a mean free path of the same order as the
lattice constant. Therefore, for the latter disorder strength the effects of the localization of
electronic eigenstates is considerable. 
Note that we have treated the disorder of the Anderson-Hubbard model exactly, 
and have solved systems having large lattices with as many as 18$^3$ sites, thereby allowing for such
localization physics to be present in our HF ground states.
 
At 1/4 filling and a disorder strength of $W/t=6$, we find a critical interaction strength of 
$U_c/t \approx 9$ leads to a metal-insulator transition, whereas a smaller disorder strength
of $W/t=2$ does not lead to any such transition for any value of $U/t$ that we have studied. 
Therefore, one concludes
that a disorder that is sufficiently strong and thereby leads to strong localization effects is
required to obtain such a transition. Of course, as always throughout our conclusions, this
is what one may conclude via a HF treatment of this model. 

\begin{itemize}
\item In order to obtain a suppression of the DOS at the Fermi level, one must allow for
the development of local magnetic moments. 
\end{itemize}

The importance of local moment formation in the disordered metallic state was made clear
in the seminal work of Milovanovic, Sachdev, and Bhatt.\cite{PhysRevLett.63.82} The 
importance of such moments in obtaining the metal-insulator transition is demonstrated
in our work.

The results from Fig.~\ref{DOS_3dSC_L18_Vdis6.0_U11.5_edensity0.5} make clear the necessity
of allowing for local moments (in a HF treatment) to form if one is to obtain a suppression
of the DOS. Further, the results from Fig.~\ref{LocalMoments}, along with
the DOS curves shown in 
Figs.~\ref{dos_conductivity_L12_Ne864_W6_U0_1_2_3}, \ref{dos_conductivity_L12_Ne864_W6_U3_4_5_6}
and \ref{dos_conductivity_L12_Ne864_W6_U6_9_11.5}, demonstrate that to obtain such a suppression
of the DOS local magnetic moments must be present. The lattice and model of disorder do not influence
whether or not such moments appear. 

We note that similar suppressions of the DOS within HF treatments of this and related models
have been found by Logan and Tusch,\cite{Tusch_Logan1993} one of us and coworkers,\cite{Fazileh_PRL}
and most recently by Shinaoka and Imada.\cite{ImadaPRL} In all of these studies a restricted
HF formulation that required magnetic moments to form along some single chosen quantization
axis was employed, again emphasizing that importance of allowing this degree of freedom. However,
whether or not ``twisted spins" that point in all directions possible are allowed\cite{chen2008} does
not seem to be required.

Due to the considerable interest in the DMFT treatment of correlated electrons, it seems appropriate
to note that theories going beyond the single-site KKR-CPA treatments of disorder within DMFT,  
such as Ref.~\cite{PhysRevB.68.125104}, also find non-Fermi-liquid behaviour. More recently,
a variant of DMFT that, like this paper, also tries to treat the disorder exactly, and
used the HF approximation in the evaluation of the off-diagonal self energy, has found\cite{Billcondmat}
zero bias anomalies similar to those shown in this paper.

\begin{itemize}
\item When such moments develop, in addition to the suppression of the DOS, one obtains a non-Drude-like 
ac conductivity. With sufficiently strong disorder and interactions, the dc conductivity is
suppressed to zero and one obtains a metal-insulator transition.
\end{itemize}

At least within our HF treatment, 
just above and just below the interaction strengths leading to the metal-insulator
transition, one finds $\sigma-\sigma_{dc} \sim \sqrt{\omega}$ and $\sigma \sim \omega^2$ power law
behaviour, the same as one finds for noninteracting electrons. Whether this is a by-product
of HF being an effective one-electron theory remains as an outstanding question. 

We note that novel ac conductivities, including results that are beyond that which we conducted, namely
that included the temperature dependencies, 
were also seen in the recent work of Kobayashi {\em et al.}\cite{kobayashi-2008} These
authors used the same HF decomposition as that used here.
However, unlike our work these authors focussed on 1/2-filled systems in two dimensions, and, in particular, 
on the properties of the novel 2d metallic phase that was predicted based on earlier 
HF work.\cite{Trivedi_PRL_2004, trivedi05a, trivedi05b} These authors do not propose specific
frequency dependencies, unlike the results we give in 
Eqs.~(\ref{metallic_powerlaw},\ref{insulating_powerlaw_fit_U9},\ref{insulating_powerlaw_fit_U11.5}),
so the potentially important effect of differing dimensionalities remains unknown.

As mentioned in the introduction, the suppression of the DOS at the Fermi energy and the associated
metal-insulator transition found in our earlier work on
lithium titanate,\cite{Fazileh_PRL} was the original motivation for this study. Since that paper
new exact diagonalization and Monte Carlo results\cite{chiesa:086401} have made clear (albeit in two
dimensions) that such physics is not an artifact of using a HF approximation. 
Subsequent HF work,\cite{ImadaPRL,ImadaJPSC}
using averages over very large numbers of complexions of disorder, has found qualitatively similar 
results, but has also managed to produce many more energy eigenvalues close to the Fermi energy,
and thereby better probe the functional form of the suppression of the DOS found in all
of these papers. 

The origin of the suppression of the DOS is not determined by the present study, although we have
placed constraints on what physics (local moments and sufficiently strong disorder) must be included 
in a model that will produce
such behaviour. In terms of existing theories, we note that for weak disorder and weak interactions
one may complete a derivation\cite{Wu} of the Altshuler and Aronov result\cite{Altshuler_Aronov_gap} for
the Anderson-Hubbard model, and one finds that one gets an increase in the DOS as $U/t$ increases
from zero. That is, one finds
\begin{equation}
\delta N(\omega) = \lambda \frac{U}{1-(UN_0(E_F))^2}~|\omega|^{1/2}
\end{equation}
where $\lambda$ is some positive constant and $N_0(E_F)$ is the DOS of the noninteracting system
at the Fermi energy.
This is consistent with the $W/t=6$ and $U/t=0,1,2$ and 3 DOS curves that we found for the 12$^3$ lattice,
as well as the $W/t=6$ and $U/t=0$ and 3 curves for the 18$^3$ lattice (even though the weak disorder
$k_f\ell_e\gg1$ assumption of the Altshuler and Aronov theory is not obeyed for our $W/t=6$ system).
For larger $U/t$ our HF numerics show that the sign of the shift of the DOS changes, in that one has a 
suppression rather than enhancement, as is clear from 
Figs.~\ref{dos_conductivity_L12_Ne864_W6_U0_1_2_3} and \ref{dos_conductivity_L12_Ne864_W6_U3_4_5_6}.
Given that we find that one must have local moments to have any suppression, even for weak disorder
(see Figs.~\ref{LocalMoments} and \ref{dos_L18_W2}), it's clear that one must extend the weak disorder
paramagnetic Fermi Altshuler and Aronov theory, which remains an outstanding problem.

However, a completely different idea was made in a recent HF and exact diagonalization 
study,\cite{ImadaPRL,ImadaJPSC}
and these authors proposed that while the Altshuler and Aronov form of
\begin{equation}\label{eq:AANofw}
N(E) \sim \mid E-E_F\mid^\delta 
\end{equation}
may be adequate away from the Fermi energy, very near the Fermi energy one instead
finds a so-called soft Hubbard gap. They showed that such physics in three dimensions gives rise to
\begin{equation}\label{eq:ImadaNofw}
N(E) \sim \exp\{-(-\gamma\log\mid E-E_F\mid)^3\} 
\end{equation}
For one-dimensional systems with strong localization effects, these authors provide independent
justification for their theory.
However, to access the energies very close to $E_F$ in both one and three dimensions, 
averages over very large numbers of complexions of disorder
were carried out.\cite{ImadaPRL,ImadaJPSC} 

Our numerics do not have such good resolution, but we have examined DOS data for an $18^3$ lattice
for $W/t=6$ and $U/t=11.5$, now averaged over three different complexions of disorder.
First, we note that with a broadening roughly 1/5 of that shown in our Fig.~\ref{DOS_L18_W6_box} we
find that the DOS indeed does indeed go to zero at $E_F$; this is consistent
with Refs.\cite{ImadaPRL,ImadaJPSC}.  Second, fitting the DOS data with these smaller
broadening factors, for $\omega/t \gtrsim 0.2$ we find that Eq.~(\ref{eq:AANofw}) fits
the data well, and we obtain $\delta = 0.51 \pm 0.01$. Therefore, we find the same
exponent as in the theory of Altshuler and Aronov.\cite{Altshuler_Aronov_gap} 
However, while our data for $\omega/t \lesssim 0.1$
is sparse and therefore not completely reliable, the functional form proposed in
Eq.~(\ref{eq:ImadaNofw}) indeed fits our DOS curves reasonably well. Our limited
numerics therefore support the soft Hubbard gap proposal\cite{ImadaPRL,ImadaJPSC} 
at energies very close to $E_F$.

Finally, we mention the relationship of our work to some experimental results on 
transition metal oxides that were mentioned in the introduction.
Sarma et al.\cite{sarma_cusp_dos} have investigated the electronic structure of LaNi$_{1-x}$Mn$_x$O$_3$. 
The conductance is proportional to $\sqrt{V}$, where $V$ is the bias voltage. As mentioned above, such
a form fits our data very well.
Kim et al.\cite{PhysRevB.71.125104} have studied the transport and optical properties of 
SrTi$_{1-x}$Ru$_x$O$_3$. Their low frequency optical conductivity data is very similar
to that which we obtained within HF. Our data also agrees with the optical conductivity found
in the disordered metallic phase of conducting polymers.\cite{PhysRevB.48.14884}
The analysis of this data, based on the localization modified
Drude model,\cite{MottKaveh} leads to the same 0.5 exponent as we obtained in our unbiased
power law fitting, the result of which is given in Eq.~(\ref{metallic_powerlaw_fit}).

\begin{acknowledgments}
We wish to thank David Johnston, Farhad Fazileh, Dariush Heidarian, Nandini Trivedi and 
Bill Atkinson for numerous helpful communications at the beginning of this work. 
This work was supported in part by the NSERC of Canada.
\end{acknowledgments}


\begin{thebibliography}{54}
\expandafter\ifx\csname natexlab\endcsname\relax\def\natexlab#1{#1}\fi
\expandafter\ifx\csname bibnamefont\endcsname\relax
  \def\bibnamefont#1{#1}\fi
\expandafter\ifx\csname bibfnamefont\endcsname\relax
  \def\bibfnamefont#1{#1}\fi
\expandafter\ifx\csname citenamefont\endcsname\relax
  \def\citenamefont#1{#1}\fi
\expandafter\ifx\csname url\endcsname\relax
  \def\url#1{\texttt{#1}}\fi
\expandafter\ifx\csname urlprefix\endcsname\relax\def\urlprefix{URL }\fi
\providecommand{\bibinfo}[2]{#2}
\providecommand{\eprint}[2][]{\url{#2}}

\bibitem[{\citenamefont{Cox}(1992)}]{CoxTMO}
\bibinfo{author}{\bibfnamefont{P.~A.} \bibnamefont{Cox}},
  \emph{\bibinfo{title}{Transition Metal Oxides: An Introduction to Their
  Electronic Structure and Properties}} (\bibinfo{publisher}{Oxford University
  Press, New York}, \bibinfo{year}{1992}).

\bibitem[{\citenamefont{Imada et~al.}(1998)\citenamefont{Imada, Fujimori, and
  Tokura}}]{Imada_MIT}
\bibinfo{author}{\bibfnamefont{M.}~\bibnamefont{Imada}},
  \bibinfo{author}{\bibfnamefont{A.}~\bibnamefont{Fujimori}}, \bibnamefont{and}
  \bibinfo{author}{\bibfnamefont{Y.}~\bibnamefont{Tokura}},
  \bibinfo{journal}{Rev. Mod. Phys.} \textbf{\bibinfo{volume}{70}},
  \bibinfo{pages}{1039} (\bibinfo{year}{1998}).

\bibitem[{\citenamefont{Lai and Gooding}(1998)}]{Lai_PRB}
\bibinfo{author}{\bibfnamefont{E.}~\bibnamefont{Lai}} \bibnamefont{and}
  \bibinfo{author}{\bibfnamefont{R.~J.} \bibnamefont{Gooding}},
  \bibinfo{journal}{Phys. Rev. B} \textbf{\bibinfo{volume}{57}},
  \bibinfo{pages}{1498} (\bibinfo{year}{1998}).

\bibitem[{\citenamefont{Kastner et~al.}(1998)\citenamefont{Kastner, Birgeneau,
  Shirane, and Endoh}}]{Kastner_RMP}
\bibinfo{author}{\bibfnamefont{M.~A.} \bibnamefont{Kastner}},
  \bibinfo{author}{\bibfnamefont{R.~J.} \bibnamefont{Birgeneau}},
  \bibinfo{author}{\bibfnamefont{G.}~\bibnamefont{Shirane}}, \bibnamefont{and}
  \bibinfo{author}{\bibfnamefont{Y.}~\bibnamefont{Endoh}},
  \bibinfo{journal}{Rev. Mod. Phys.} \textbf{\bibinfo{volume}{70}},
  \bibinfo{pages}{897} (\bibinfo{year}{1998}).

\bibitem[{\citenamefont{Lambert et~al.}(1990)\citenamefont{Lambert, Edwards,
  and Harrison}}]{Lambert90}
\bibinfo{author}{\bibfnamefont{P.~M.} \bibnamefont{Lambert}},
  \bibinfo{author}{\bibfnamefont{P.~P.} \bibnamefont{Edwards}},
  \bibnamefont{and} \bibinfo{author}{\bibfnamefont{M.~R.}
  \bibnamefont{Harrison}}, \bibinfo{journal}{J. Sol. St. Chem.}
  \textbf{\bibinfo{volume}{89}}, \bibinfo{pages}{345} (\bibinfo{year}{1990}).

\bibitem[{\citenamefont{Johnston}(1976)}]{Johnston76I}
\bibinfo{author}{\bibfnamefont{D.~C.} \bibnamefont{Johnston}},
  \bibinfo{journal}{J. Low Temp. Phys.} \textbf{\bibinfo{volume}{25}},
  \bibinfo{pages}{145} (\bibinfo{year}{1976}).

\bibitem[{\citenamefont{McCallum et~al.}(1976)\citenamefont{McCallum, Johnston,
  Luengo, and Maple}}]{Johnston76II}
\bibinfo{author}{\bibfnamefont{R.~W.} \bibnamefont{McCallum}},
  \bibinfo{author}{\bibfnamefont{D.~C.} \bibnamefont{Johnston}},
  \bibinfo{author}{\bibfnamefont{C.~A.} \bibnamefont{Luengo}},
  \bibnamefont{and} \bibinfo{author}{\bibfnamefont{M.~B.} \bibnamefont{Maple}},
  \bibinfo{journal}{J. Low Temp. Phys.} \textbf{\bibinfo{volume}{25}},
  \bibinfo{pages}{177} (\bibinfo{year}{1976}).

\bibitem[{\citenamefont{M{\"u}ller}(1996)}]{Mueller96}
\bibinfo{author}{\bibfnamefont{K.~A.} \bibnamefont{M{\"u}ller}}, in
  \emph{\bibinfo{booktitle}{{Proceedings of the 10th Anniversary {HTS} Workshop
  on Physics, Materials, and Applications}}}, edited by
  \bibinfo{editor}{\bibfnamefont{B.}~\bibnamefont{Batlogg}}
  \bibnamefont{et~al.} (\bibinfo{publisher}{World Scientific, Singapore},
  \bibinfo{year}{1996}), p.~\bibinfo{pages}{1}.

\bibitem[{\citenamefont{Fazileh et~al.}(2004)\citenamefont{Fazileh, Gooding,
  and Johnston}}]{Fazileh_PRB}
\bibinfo{author}{\bibfnamefont{F.}~\bibnamefont{Fazileh}},
  \bibinfo{author}{\bibfnamefont{R.~J.} \bibnamefont{Gooding}},
  \bibnamefont{and} \bibinfo{author}{\bibfnamefont{D.~C.}
  \bibnamefont{Johnston}}, \bibinfo{journal}{Phys. Rev. B}
  \textbf{\bibinfo{volume}{69}}, \bibinfo{pages}{104503}
  (\bibinfo{year}{2004}).

\bibitem[{\citenamefont{Fazileh et~al.}(2006)\citenamefont{Fazileh, Gooding,
  Atkinson, and Johnston}}]{Fazileh_PRL}
\bibinfo{author}{\bibfnamefont{F.}~\bibnamefont{Fazileh}},
  \bibinfo{author}{\bibfnamefont{R.~J.} \bibnamefont{Gooding}},
  \bibinfo{author}{\bibfnamefont{W.~A.} \bibnamefont{Atkinson}},
  \bibnamefont{and} \bibinfo{author}{\bibfnamefont{D.~C.}
  \bibnamefont{Johnston}}, \bibinfo{journal}{Phys. Rev. Lett.}
  \textbf{\bibinfo{volume}{96}}, \bibinfo{pages}{046410}
  (\bibinfo{year}{2006}).

\bibitem[{\citenamefont{Sarma et~al.}(1998)\citenamefont{Sarma, Chainani,
  Krishnakumar, Vescovo, Carbone, Eberhardt, Rader, Jung, Hellwig, Gudat
  et~al.}}]{sarma_cusp_dos}
\bibinfo{author}{\bibfnamefont{D.~D.} \bibnamefont{Sarma}},
  \bibinfo{author}{\bibfnamefont{A.}~\bibnamefont{Chainani}},
  \bibinfo{author}{\bibfnamefont{S.~R.} \bibnamefont{Krishnakumar}},
  \bibinfo{author}{\bibfnamefont{E.}~\bibnamefont{Vescovo}},
  \bibinfo{author}{\bibfnamefont{C.}~\bibnamefont{Carbone}},
  \bibinfo{author}{\bibfnamefont{W.}~\bibnamefont{Eberhardt}},
  \bibinfo{author}{\bibfnamefont{O.}~\bibnamefont{Rader}},
  \bibinfo{author}{\bibfnamefont{C.}~\bibnamefont{Jung}},
  \bibinfo{author}{\bibfnamefont{C.}~\bibnamefont{Hellwig}},
  \bibinfo{author}{\bibfnamefont{W.}~\bibnamefont{Gudat}},
  \bibnamefont{et~al.}, \bibinfo{journal}{Phys. Rev. Lett.}
  \textbf{\bibinfo{volume}{80}}, \bibinfo{pages}{4004} (\bibinfo{year}{1998}).

\bibitem[{\citenamefont{Altshuler and Aronov}(1979)}]{Altshuler_Aronov_gap}
\bibinfo{author}{\bibfnamefont{B.~L.} \bibnamefont{Altshuler}}
  \bibnamefont{and} \bibinfo{author}{\bibfnamefont{A.~G.}
  \bibnamefont{Aronov}}, \bibinfo{journal}{Solid State Commun.}
  \textbf{\bibinfo{volume}{30}}, \bibinfo{pages}{115} (\bibinfo{year}{1979}).

\bibitem[{\citenamefont{Kim et~al.}(2005)\citenamefont{Kim, Lee, Noh, Lee, and
  Char}}]{PhysRevB.71.125104}
\bibinfo{author}{\bibfnamefont{K.~W.} \bibnamefont{Kim}},
  \bibinfo{author}{\bibfnamefont{J.~S.} \bibnamefont{Lee}},
  \bibinfo{author}{\bibfnamefont{T.~W.} \bibnamefont{Noh}},
  \bibinfo{author}{\bibfnamefont{S.~R.} \bibnamefont{Lee}}, \bibnamefont{and}
  \bibinfo{author}{\bibfnamefont{K.}~\bibnamefont{Char}},
  \bibinfo{journal}{Phys. Rev. B} \textbf{\bibinfo{volume}{71}},
  \bibinfo{pages}{125104} (\bibinfo{year}{2005}).

\bibitem[{\citenamefont{Grussbach and Schreiber}(1995)}]{muC}
\bibinfo{author}{\bibfnamefont{H.}~\bibnamefont{Grussbach}} \bibnamefont{and}
  \bibinfo{author}{\bibfnamefont{M.}~\bibnamefont{Schreiber}},
  \bibinfo{journal}{Phys. Rev. B} \textbf{\bibinfo{volume}{51}},
  \bibinfo{pages}{663} (\bibinfo{year}{1995}).

\bibitem[{\citenamefont{Paris et~al.}(2007)\citenamefont{Paris, Baldwin, and
  Scalettar}}]{paris:165113}
\bibinfo{author}{\bibfnamefont{N.}~\bibnamefont{Paris}},
  \bibinfo{author}{\bibfnamefont{A.}~\bibnamefont{Baldwin}}, \bibnamefont{and}
  \bibinfo{author}{\bibfnamefont{R.~T.} \bibnamefont{Scalettar}},
  \bibinfo{journal}{Phys. Rev. B} \textbf{\bibinfo{volume}{75}},
  \bibinfo{eid}{165113} (\bibinfo{year}{2007}).

\bibitem[{\citenamefont{Benenti et~al.}(1999)\citenamefont{Benenti, Waintal,
  and Pichard}}]{PhysRevLett.83.1826}
\bibinfo{author}{\bibfnamefont{G.}~\bibnamefont{Benenti}},
  \bibinfo{author}{\bibfnamefont{X.}~\bibnamefont{Waintal}}, \bibnamefont{and}
  \bibinfo{author}{\bibfnamefont{J.-L.} \bibnamefont{Pichard}},
  \bibinfo{journal}{Phys. Rev. Lett.} \textbf{\bibinfo{volume}{83}},
  \bibinfo{pages}{1826} (\bibinfo{year}{1999}).

\bibitem[{\citenamefont{Berkovits et~al.}(2001)\citenamefont{Berkovits,
  Kantelhardt, Avishai, Havlin, and Bunde}}]{PhysRevB.63.085102}
\bibinfo{author}{\bibfnamefont{R.}~\bibnamefont{Berkovits}},
  \bibinfo{author}{\bibfnamefont{J.~W.} \bibnamefont{Kantelhardt}},
  \bibinfo{author}{\bibfnamefont{Y.}~\bibnamefont{Avishai}},
  \bibinfo{author}{\bibfnamefont{S.}~\bibnamefont{Havlin}}, \bibnamefont{and}
  \bibinfo{author}{\bibfnamefont{A.}~\bibnamefont{Bunde}},
  \bibinfo{journal}{Phys. Rev. B} \textbf{\bibinfo{volume}{63}},
  \bibinfo{pages}{085102} (\bibinfo{year}{2001}).

\bibitem[{\citenamefont{Kotlyar and Das~Sarma}(2001)}]{PhysRevLett.86.2388}
\bibinfo{author}{\bibfnamefont{R.}~\bibnamefont{Kotlyar}} \bibnamefont{and}
  \bibinfo{author}{\bibfnamefont{S.}~\bibnamefont{Das~Sarma}},
  \bibinfo{journal}{Phys. Rev. Lett.} \textbf{\bibinfo{volume}{86}},
  \bibinfo{pages}{2388} (\bibinfo{year}{2001}).

\bibitem[{\citenamefont{Vasseur and Weinmann}(2004)}]{epjb.42.279}
\bibinfo{author}{\bibfnamefont{G.}~\bibnamefont{Vasseur}} \bibnamefont{and}
  \bibinfo{author}{\bibfnamefont{D.}~\bibnamefont{Weinmann}},
  \bibinfo{journal}{Eur. Phys. J. B} \textbf{\bibinfo{volume}{42}},
  \bibinfo{pages}{279} (\bibinfo{year}{2004}).

\bibitem[{\citenamefont{Shinaoka and Imada}(2009{\natexlab{a}})}]{ImadaPRL}
\bibinfo{author}{\bibfnamefont{H.}~\bibnamefont{Shinaoka}} \bibnamefont{and}
  \bibinfo{author}{\bibfnamefont{M.}~\bibnamefont{Imada}},
  \bibinfo{journal}{Phys. Rev. Lett.} \textbf{\bibinfo{volume}{102}},
  \bibinfo{pages}{016404} (\bibinfo{year}{2009}{\natexlab{a}}).

\bibitem[{\citenamefont{Ulmke and Scalettar}(1997)}]{PhysRevB.55.4149}
\bibinfo{author}{\bibfnamefont{M.}~\bibnamefont{Ulmke}} \bibnamefont{and}
  \bibinfo{author}{\bibfnamefont{R.~T.} \bibnamefont{Scalettar}},
  \bibinfo{journal}{Phys. Rev. B} \textbf{\bibinfo{volume}{55}},
  \bibinfo{pages}{4149} (\bibinfo{year}{1997}).

\bibitem[{\citenamefont{Denteneer et~al.}(1999)\citenamefont{Denteneer,
  Scalettar, and Trivedi}}]{PhysRevLett.83.4610}
\bibinfo{author}{\bibfnamefont{P.~J.~H.} \bibnamefont{Denteneer}},
  \bibinfo{author}{\bibfnamefont{R.~T.} \bibnamefont{Scalettar}},
  \bibnamefont{and} \bibinfo{author}{\bibfnamefont{N.}~\bibnamefont{Trivedi}},
  \bibinfo{journal}{Phys. Rev. Lett.} \textbf{\bibinfo{volume}{83}},
  \bibinfo{pages}{4610} (\bibinfo{year}{1999}).

\bibitem[{\citenamefont{Caldara et~al.}(2000)\citenamefont{Caldara, Srinivasan,
  and Shepelyansky}}]{PhysRevB.62.10680}
\bibinfo{author}{\bibfnamefont{G.}~\bibnamefont{Caldara}},
  \bibinfo{author}{\bibfnamefont{B.}~\bibnamefont{Srinivasan}},
  \bibnamefont{and} \bibinfo{author}{\bibfnamefont{D.~L.}
  \bibnamefont{Shepelyansky}}, \bibinfo{journal}{Phys. Rev. B}
  \textbf{\bibinfo{volume}{62}}, \bibinfo{pages}{10680} (\bibinfo{year}{2000}).

\bibitem[{\citenamefont{Enjalran et~al.}(2001)\citenamefont{Enjalran, H\'ebert,
  Batrouni, Scalettar, and Zhang}}]{PhysRevB.64.184402}
\bibinfo{author}{\bibfnamefont{M.}~\bibnamefont{Enjalran}},
  \bibinfo{author}{\bibfnamefont{F.}~\bibnamefont{H\'ebert}},
  \bibinfo{author}{\bibfnamefont{G.~G.} \bibnamefont{Batrouni}},
  \bibinfo{author}{\bibfnamefont{R.~T.} \bibnamefont{Scalettar}},
  \bibnamefont{and} \bibinfo{author}{\bibfnamefont{S.}~\bibnamefont{Zhang}},
  \bibinfo{journal}{Phys. Rev. B} \textbf{\bibinfo{volume}{64}},
  \bibinfo{pages}{184402} (\bibinfo{year}{2001}).

\bibitem[{\citenamefont{Denteneer et~al.}(2001)\citenamefont{Denteneer,
  Scalettar, and Trivedi}}]{PhysRevLett.87.146401}
\bibinfo{author}{\bibfnamefont{P.~J.~H.} \bibnamefont{Denteneer}},
  \bibinfo{author}{\bibfnamefont{R.~T.} \bibnamefont{Scalettar}},
  \bibnamefont{and} \bibinfo{author}{\bibfnamefont{N.}~\bibnamefont{Trivedi}},
  \bibinfo{journal}{Phys. Rev. Lett.} \textbf{\bibinfo{volume}{87}},
  \bibinfo{pages}{146401} (\bibinfo{year}{2001}).

\bibitem[{\citenamefont{Srinivasan et~al.}(2003)\citenamefont{Srinivasan,
  Benenti, and Shepelyansky}}]{PhysRevB.67.205112}
\bibinfo{author}{\bibfnamefont{B.}~\bibnamefont{Srinivasan}},
  \bibinfo{author}{\bibfnamefont{G.}~\bibnamefont{Benenti}}, \bibnamefont{and}
  \bibinfo{author}{\bibfnamefont{D.~L.} \bibnamefont{Shepelyansky}},
  \bibinfo{journal}{Phys. Rev. B} \textbf{\bibinfo{volume}{67}},
  \bibinfo{pages}{205112} (\bibinfo{year}{2003}).

\bibitem[{\citenamefont{Chakraborty et~al.}(2007)\citenamefont{Chakraborty,
  Denteneer, and Scalettar}}]{chakraborty:125117}
\bibinfo{author}{\bibfnamefont{P.~B.} \bibnamefont{Chakraborty}},
  \bibinfo{author}{\bibfnamefont{P.~J.~H.} \bibnamefont{Denteneer}},
  \bibnamefont{and} \bibinfo{author}{\bibfnamefont{R.~T.}
  \bibnamefont{Scalettar}}, \bibinfo{journal}{Phys. Rev. B}
  \textbf{\bibinfo{volume}{75}}, \bibinfo{eid}{125117} (\bibinfo{year}{2007}).

\bibitem[{\citenamefont{Chiesa et~al.}(2008)\citenamefont{Chiesa, Chakraborty,
  Pickett, and Scalettar}}]{chiesa:086401}
\bibinfo{author}{\bibfnamefont{S.}~\bibnamefont{Chiesa}},
  \bibinfo{author}{\bibfnamefont{P.~B.} \bibnamefont{Chakraborty}},
  \bibinfo{author}{\bibfnamefont{W.~E.} \bibnamefont{Pickett}},
  \bibnamefont{and} \bibinfo{author}{\bibfnamefont{R.~T.}
  \bibnamefont{Scalettar}}, \bibinfo{journal}{Phys. Rev. Lett.}
  \textbf{\bibinfo{volume}{101}}, \bibinfo{eid}{086401} (\bibinfo{year}{2008}).

\bibitem[{\citenamefont{Otsuka and Hatsugai}(2000)}]{Otsuka00}
\bibinfo{author}{\bibfnamefont{Y.}~\bibnamefont{Otsuka}} \bibnamefont{and}
  \bibinfo{author}{\bibfnamefont{Y.}~\bibnamefont{Hatsugai}},
  \bibinfo{journal}{J. Phys. Chem.} \textbf{\bibinfo{volume}{12}},
  \bibinfo{pages}{9317} (\bibinfo{year}{2000}).

\bibitem[{\citenamefont{Chen et~al.}(2008)\citenamefont{Chen, Farhoodfar,
  McIntosh, Gooding, and Leung}}]{chen2008}
\bibinfo{author}{\bibfnamefont{X.}~\bibnamefont{Chen}},
  \bibinfo{author}{\bibfnamefont{A.}~\bibnamefont{Farhoodfar}},
  \bibinfo{author}{\bibfnamefont{T.}~\bibnamefont{McIntosh}},
  \bibinfo{author}{\bibfnamefont{R.~J.} \bibnamefont{Gooding}},
  \bibnamefont{and} \bibinfo{author}{\bibfnamefont{P.~W.} \bibnamefont{Leung}},
  \bibinfo{journal}{J. Phys.: Condens. Matter} \textbf{\bibinfo{volume}{20}},
  \bibinfo{pages}{345211} (\bibinfo{year}{2008}).

\bibitem[{\citenamefont{Dasgupta and Halley}(1993)}]{PhysRevB.47.1126}
\bibinfo{author}{\bibfnamefont{C.}~\bibnamefont{Dasgupta}} \bibnamefont{and}
  \bibinfo{author}{\bibfnamefont{J.~W.} \bibnamefont{Halley}},
  \bibinfo{journal}{Phys. Rev. B} \textbf{\bibinfo{volume}{47}},
  \bibinfo{pages}{1126} (\bibinfo{year}{1993}).

\bibitem[{\citenamefont{Heidarian and Trivedi}(2004)}]{Trivedi_PRL_2004}
\bibinfo{author}{\bibfnamefont{D.}~\bibnamefont{Heidarian}} \bibnamefont{and}
  \bibinfo{author}{\bibfnamefont{N.}~\bibnamefont{Trivedi}},
  \bibinfo{journal}{Phys. Rev. Lett.} \textbf{\bibinfo{volume}{93}},
  \bibinfo{pages}{126401} (\bibinfo{year}{2004}).

\bibitem[{\citenamefont{Trivedi and Heidarian}(2005)}]{trivedi05a}
\bibinfo{author}{\bibfnamefont{N.}~\bibnamefont{Trivedi}} \bibnamefont{and}
  \bibinfo{author}{\bibfnamefont{D.}~\bibnamefont{Heidarian}},
  \bibinfo{journal}{Prog. Theor. Phys. Suppl.} \textbf{\bibinfo{volume}{160}},
  \bibinfo{pages}{296} (\bibinfo{year}{2005}).

\bibitem[{\citenamefont{Trivedi et~al.}(2005)\citenamefont{Trivedi, Denteneer,
  Heidarian, and Scalettar}}]{trivedi05b}
\bibinfo{author}{\bibfnamefont{N.}~\bibnamefont{Trivedi}},
  \bibinfo{author}{\bibfnamefont{P.~J.~H.} \bibnamefont{Denteneer}},
  \bibinfo{author}{\bibfnamefont{D.}~\bibnamefont{Heidarian}},
  \bibnamefont{and} \bibinfo{author}{\bibfnamefont{R.~T.}
  \bibnamefont{Scalettar}}, \bibinfo{journal}{Pramana}
  \textbf{\bibinfo{volume}{64}}, \bibinfo{pages}{1051} (\bibinfo{year}{2005}).

\bibitem[{\citenamefont{Kobayashi et~al.}()\citenamefont{Kobayashi, Lee, and
  Trivedi}}]{kobayashi-2008}
\bibinfo{author}{\bibfnamefont{K.}~\bibnamefont{Kobayashi}},
  \bibinfo{author}{\bibfnamefont{B.}~\bibnamefont{Lee}}, \bibnamefont{and}
  \bibinfo{author}{\bibfnamefont{N.}~\bibnamefont{Trivedi}},
  \bibinfo{note}{arXiv.org:0807.3372}.

\bibitem[{\citenamefont{Tusch and Logan}(1993)}]{Tusch_Logan1993}
\bibinfo{author}{\bibfnamefont{M.~A.} \bibnamefont{Tusch}} \bibnamefont{and}
  \bibinfo{author}{\bibfnamefont{D.~E.} \bibnamefont{Logan}},
  \bibinfo{journal}{Phys. Rev. B} \textbf{\bibinfo{volume}{48}},
  \bibinfo{pages}{14843} (\bibinfo{year}{1993}).

\bibitem[{\citenamefont{D\"{u}cker et~al.}(1999)\citenamefont{D\"{u}cker, von
  Niessen, Koslowski, Tusch, and Logan}}]{duecker99}
\bibinfo{author}{\bibfnamefont{H.}~\bibnamefont{D\"{u}cker}},
  \bibinfo{author}{\bibfnamefont{W.}~\bibnamefont{von Niessen}},
  \bibinfo{author}{\bibfnamefont{T.}~\bibnamefont{Koslowski}},
  \bibinfo{author}{\bibfnamefont{M.~A.} \bibnamefont{Tusch}}, \bibnamefont{and}
  \bibinfo{author}{\bibfnamefont{D.~E.} \bibnamefont{Logan}},
  \bibinfo{journal}{Phys. Rev. B} \textbf{\bibinfo{volume}{59}},
  \bibinfo{pages}{871} (\bibinfo{year}{1999}).

\bibitem[{\citenamefont{Mirlin}(2000)}]{mirlin}
\bibinfo{author}{\bibfnamefont{A.~D.} \bibnamefont{Mirlin}},
  \bibinfo{journal}{Phys. Rep.} \textbf{\bibinfo{volume}{326}},
  \bibinfo{pages}{259} (\bibinfo{year}{2000}).

\bibitem[{\citenamefont{Milovanovi{\'c}
  et~al.}(1989)\citenamefont{Milovanovi{\'c}, Sachdev, and
  Bhatt}}]{PhysRevLett.63.82}
\bibinfo{author}{\bibfnamefont{M.}~\bibnamefont{Milovanovi{\'c}}},
  \bibinfo{author}{\bibfnamefont{S.}~\bibnamefont{Sachdev}}, \bibnamefont{and}
  \bibinfo{author}{\bibfnamefont{R.~N.} \bibnamefont{Bhatt}},
  \bibinfo{journal}{Phys. Rev. Lett.} \textbf{\bibinfo{volume}{63}},
  \bibinfo{pages}{82} (\bibinfo{year}{1989}).

\bibitem[{\citenamefont{Fazekas}(1999)}]{Fazekas}
\bibinfo{author}{\bibfnamefont{P.}~\bibnamefont{Fazekas}},
  \emph{\bibinfo{title}{Lecture Notes on Electron Correlation and Magnetism}}
  (\bibinfo{publisher}{World Scientific, Singapore}, \bibinfo{year}{1999}).

\bibitem[{\citenamefont{Mahan}(2000)}]{Mahanbook}
\bibinfo{author}{\bibfnamefont{G.~D.} \bibnamefont{Mahan}},
  \emph{\bibinfo{title}{Many Particle Physics}} (\bibinfo{publisher}{Kluwer
  Academic, New York}, \bibinfo{year}{2000}).

\bibitem[{\citenamefont{Mott}(1990)}]{Mott_MIT}
\bibinfo{author}{\bibfnamefont{N.}~\bibnamefont{Mott}},
  \emph{\bibinfo{title}{Metal-Insulator Transitions}}
  (\bibinfo{publisher}{Taylor and Francis, New York}, \bibinfo{year}{1990}).

\bibitem[{\citenamefont{Fazileh}(2005)}]{Farhad_Thesis}
\bibinfo{author}{\bibfnamefont{F.}~\bibnamefont{Fazileh}}, Ph.D. thesis,
  \bibinfo{school}{Queen's University} (\bibinfo{year}{2005}).

\bibitem[{\citenamefont{Edwards and Anderson}(1975)}]{EA}
\bibinfo{author}{\bibfnamefont{S.~F.} \bibnamefont{Edwards}} \bibnamefont{and}
  \bibinfo{author}{\bibfnamefont{P.~W.} \bibnamefont{Anderson}},
  \bibinfo{journal}{J. Phys. F} \textbf{\bibinfo{volume}{5}},
  \bibinfo{pages}{965} (\bibinfo{year}{1975}).

\bibitem[{\citenamefont{Xue and Lee}(1988)}]{PhysRevB.38.9093}
\bibinfo{author}{\bibfnamefont{W.}~\bibnamefont{Xue}} \bibnamefont{and}
  \bibinfo{author}{\bibfnamefont{P.~A.} \bibnamefont{Lee}},
  \bibinfo{journal}{Phys. Rev. B} \textbf{\bibinfo{volume}{38}},
  \bibinfo{pages}{9093} (\bibinfo{year}{1988}).

\bibitem[{\citenamefont{Mott. and Kaveh}(1985)}]{MottKaveh}
\bibinfo{author}{\bibfnamefont{N.~F.} \bibnamefont{Mott.}} \bibnamefont{and}
  \bibinfo{author}{\bibfnamefont{M.}~\bibnamefont{Kaveh}},
  \bibinfo{journal}{Adv. Phys.} \textbf{\bibinfo{volume}{34}},
  \bibinfo{pages}{329} (\bibinfo{year}{1985}).

\bibitem[{\citenamefont{Lee et~al.}(1993)\citenamefont{Lee, Heeger, and
  Cao}}]{PhysRevB.48.14884}
\bibinfo{author}{\bibfnamefont{K.}~\bibnamefont{Lee}},
  \bibinfo{author}{\bibfnamefont{A.~J.} \bibnamefont{Heeger}},
  \bibnamefont{and} \bibinfo{author}{\bibfnamefont{Y.}~\bibnamefont{Cao}},
  \bibinfo{journal}{Phys. Rev. B} \textbf{\bibinfo{volume}{48}},
  \bibinfo{pages}{14884} (\bibinfo{year}{1993}).

\bibitem[{\citenamefont{Kroha}(1990)}]{Kroha1990231}
\bibinfo{author}{\bibfnamefont{J.}~\bibnamefont{Kroha}},
  \bibinfo{journal}{Physica A} \textbf{\bibinfo{volume}{167}},
  \bibinfo{pages}{231 } (\bibinfo{year}{1990}).

\bibitem[{\citenamefont{Song et~al.}(2008)\citenamefont{Song, Wortis, and
  Atkinson}}]{Bill2008}
\bibinfo{author}{\bibfnamefont{Y.~S.} \bibnamefont{Song}},
  \bibinfo{author}{\bibfnamefont{R.}~\bibnamefont{Wortis}}, \bibnamefont{and}
  \bibinfo{author}{\bibfnamefont{W.~A.} \bibnamefont{Atkinson}},
  \bibinfo{journal}{Phys. Rev. B} \textbf{\bibinfo{volume}{77}},
  \bibinfo{pages}{054202} (\bibinfo{year}{2008}).

\bibitem[{\citenamefont{Henseler et~al.}(2008)\citenamefont{Henseler, Kroha,
  and Shapiro}}]{Kroha2008}
\bibinfo{author}{\bibfnamefont{P.}~\bibnamefont{Henseler}},
  \bibinfo{author}{\bibfnamefont{J.}~\bibnamefont{Kroha}}, \bibnamefont{and}
  \bibinfo{author}{\bibfnamefont{B.}~\bibnamefont{Shapiro}},
  \bibinfo{journal}{Phys. Rev. B} \textbf{\bibinfo{volume}{78}},
  \bibinfo{pages}{235116} (\bibinfo{year}{2008}).

\bibitem[{\citenamefont{Aguiar et~al.}(2003)\citenamefont{Aguiar, Miranda, and
  Dobrosavljevi{\' c}}}]{PhysRevB.68.125104}
\bibinfo{author}{\bibfnamefont{M.~C.~O.} \bibnamefont{Aguiar}},
  \bibinfo{author}{\bibfnamefont{E.}~\bibnamefont{Miranda}}, \bibnamefont{and}
  \bibinfo{author}{\bibfnamefont{V.}~\bibnamefont{Dobrosavljevi{\' c}}},
  \bibinfo{journal}{Phys. Rev. B} \textbf{\bibinfo{volume}{68}},
  \bibinfo{pages}{125104} (\bibinfo{year}{2003}).

\bibitem[{\citenamefont{Song et~al.}()\citenamefont{Song, Bulut, Wortis, and
  Atkinson}}]{Billcondmat}
\bibinfo{author}{\bibfnamefont{Y.~S.} \bibnamefont{Song}},
  \bibinfo{author}{\bibfnamefont{S.}~\bibnamefont{Bulut}},
  \bibinfo{author}{\bibfnamefont{R.}~\bibnamefont{Wortis}}, \bibnamefont{and}
  \bibinfo{author}{\bibfnamefont{W.~A.} \bibnamefont{Atkinson}},
  \bibinfo{note}{arXiv:0808.3356}.

\bibitem[{\citenamefont{Shinaoka and Imada}(2009{\natexlab{b}})}]{ImadaJPSC}
\bibinfo{author}{\bibfnamefont{H.}~\bibnamefont{Shinaoka}} \bibnamefont{and}
  \bibinfo{author}{\bibfnamefont{M.}~\bibnamefont{Imada}}, \bibinfo{journal}{J.
  Phys. Soc. Jpn.}  (\bibinfo{year}{2009}{\natexlab{b}}), \bibinfo{note}{to be
  published}.

\bibitem[{Wu()}]{Wu}
\bibinfo{note}{A. Wu and R.J. Gooding (unpublished)}.

\end{thebibliography}

\end{document}